%% file: ms.tex
\documentclass[acmsmall]{acmart}

\AtBeginDocument{%
  \providecommand\BibTeX{{%
    \normalfont B\kern-0.5em{\scshape i\kern-0.25em b}\kern-0.8em\TeX}}}

\usepackage{devanagari}
\usepackage{caption}
\usepackage{subcaption}
\usepackage{graphicx}
\usepackage{hyperref} 

\usepackage{url}

\usepackage{listings}

\usepackage{fancyhdr,graphicx}

\usepackage{amsmath}
\usepackage{float}
\usepackage[export]{adjustbox}
\DeclareUnicodeCharacter{2212}{\textendash}

\graphicspath{{Pictures/}} 

\providecommand{\keys}[1]
{
  \small	
  \textbf{\textit{Keywords---}} #1
}

\begin{document}

\rhead{\thepage}

\title{Influence of NaMo App on Twitter}
\lhead{Influence of NaMo App on Twitter}

\author{Shreya Sharma}
\authornote{Both authors contributed equally to this research.}
\email{shreyasa@iitk.ac.in}
\affiliation{%
  \institution{Indian Institute of Technology}
  \country{Kanpur}
}

\author{Samiya Caur}
\authornotemark[1]
\email{samiyac@iiitd.ac.in}
\affiliation{%
  \institution{Indraprastha Institute of Information Technology}
  \country{Delhi}
  }

\author{Hitkul}

\email{hitkuli@iiitd.ac.in}
\affiliation{%
  \institution{Indraprastha Institute of Information Technology}
  \country{Delhi}
  }
  
  \author{Ponnurangam Kumaraguru}

\email{pk.guru@iiit.ac.in}
\affiliation{%
  \institution{International Institute of Information Technology}
  \country{Hyderabad}
  }

\begin{abstract}
Social media plays a crucial role in today’s society. It results in paradigm changes in how people relate and communicate, convey and exchange ideas. Moreover, social media has evolved into critical knowledge networks for consumers and also affects decision-making. In elections, social media became an integral part of political campaigning to reach a greater audience and gather more support. The 2019 Lok Sabha election saw a massive spike in the usage of online social media platforms such as Twitter, Facebook, and WhatsApp; with every major political party launching its own organized social media campaigns by investing vast amounts of money. Without limiting mainstream media, almost all Indian political leaders started using social media, especially Facebook and Twitter, to express themselves. In 2014, Bhartiya Janta Party (BJP) took one step ahead in organizing the campaign by launching its app - NaMo App.\\
We focus our research on Twitter and NaMo App during the 2019 Lok Sabha elections and Citizenship Amendment Act (CAA) protests. Twitter is one such platform where every individual can express their views and is not biased. In contrast, NaMo App is one of the first apps centered around a specific political party. It acted as a digital medium
for Bhartiya Janta Party (BJP) for organizing the political campaign to make people’s opinion in their favor. This research aims to characterize the role of the NaMo App in a more traditional network as Twitter in shaping political discourse and studies the existence of an online echo chamber on Twitter. We began by analyzing the amount and type of content shared using the NaMo App on Twitter. We performed content and network analysis for the existence of the echo chamber. We also applied the Hawkes process to see the influence that NaMo App has on Twitter. Through this research, we can conclude that the users who share content using NaMo App, may be part of an online echo chamber and are likely to be BJP workers. We show the reach and influence that the NaMo App has on Twitter is significantly less, indicating its inability to break through the diverse audience and change the narrative on Twitter. 

\end{abstract}

\maketitle

\keys{Social Media Analysis · Computational Social Science · Social Computing · Twitter · NaMo App}

\input{Chapters/Chapter1}

\input{Chapters/Chapter2}

\input{Chapters/Chapter3}
\input{Chapters/Chapter4}

\input{Chapters/Chapter5}
\input{Chapters/Chapter6}
\input{Chapters/Chapter7}

\input{Chapters/Chapter8}
\input{Chapters/Chapter9}

\input{Chapters/Chapter10}

\nocite{*}
\label{Bibliography}


\bibliographystyle{acm} 

\bibliography{Bibliography} 

\end{document}

%% file: Chapters/Chapter1.tex

\section{\textbf{Introduction}}

\subsection{\textbf{Social Media}}
In today's digital era, Social Media plays a very prominent role. It has changed how individuals interact, communicate and influence one another and how information is gathered and disseminated. Through social media, a user can get minute-by-minute information, details about events and schedules. A user can interact and dialogue with other individual. The amount of content generated by traditional media is much lower than the content generated by social media sites, and its ability to spread by word of mouth is significantly less than that of the power of the network of numerous social media platforms. According to a report \cite{36}, the top ten number of users for most social media sites were from India. In India, for example, WhatsApp has 200 million users, Twitter has 7.65 million users, Facebook has 300 million users, and 41 million people watch YouTube every month. 

\subsection{\textbf{The 2019 Lok Sabha Election}}
The 2019 Lok sabha election was unique in many ways. National Democratic Alliance (NDA) came into power two times consecutively (2014 and 2019) with a full majority. It was undoubtedly the most polarizing election result in the history of Indian elections because BJP-led NDA alone won 352 out of the 543 seats \cite{39}. The single biggest factor explaining the BJP’s phenomenal performance was sharp polarization over the issue of Hindu nationalism \cite{58}. More than 15 million first-time voters aged between 18 and 19 were added to the nearly 900 million electorates in the 2019 General Election \cite{38}. According to Dale et al. \cite{40}, nearly half of India's 900 million registered voters were women. Not only did the 2019 election saw the greatest number of woman candidates in the elections, but also the highest number of women ever to win. The 2019 elections also witnessed a record turnout of 67.11\% - highest since independence \cite{42}.\\
The 2019 elections saw a huge Modi wave. Opposition parties also contributed by focusing on attacking Prime Minister Modi.
Having to defend its record in power, the Prime Minister's campaign in 2019 was distinctly different. BJP shifted its focus from development in 2014 to national security by adding nationalism and patriotism to it \cite{43}. BJP manifesto comprise of strong hindutva issues, whereas as opposition raised issues of employment, economic hardships, agrarian distress, economic growth, and high-end corruption \cite{43}.

\subsection{\textbf{Impact of Social Media on Politics}}
In politics, Social Media is used extensively to organize election campaigns. Political parties use social media in an attempt to reach more voters in the new media society. They are not limited to using posters and banners to reach out to voters. Politicians now convey their messages through infomercials, advertising, blog posts, and thousands of tweets and gauge their communication by watching direct reactions to their acts on social media sites. \\
The 2008 presidential election in the United States is widely seen as the first election in which social media sites played a big part. Barack Obama made extensive use of digital media such as Facebook in remarkable ways to reach younger audiences \cite{1}. This political campaigning strategy revolutionized political campaigning processes all over the world. \\
In keeping with the current growing trend, Indian political parties have also adopted social platforms for political communication to directly connect with their supporters by investing heavily into expanding their Information and Technology (IT) cells or social media campaigns. Following Prime Minister Narendra Modi's electoral success with Twitter and Facebook in 2014 \cite{2}, politicians from different states started creating Twitter profiles. Political parties like Bharatiya Janata Party (BJP), Aam Aadmi Party (AAP), and Indian National Congress (INC) seek suggestions from the Social Media users for their election manifestoes, but people's participation was limited \cite{37}. The usage of these platforms has increased politicians connectivity with the young Indian voters \cite{3}. \\
In the 2019 Lok Sabha elections, the use of such platforms reached an all-time high, with every major political party releasing their own coordinated social media campaigns that had a cumulative cost of 586.7 milions \cite{4}. The 2019 election Lok Sabha election had 130 million first-time voters, out of which more than 15 million voters between 18 to 19 years of age \cite{37}. Out of all social media platforms, Twitter was the most popular.

\subsection{\textbf{Echo Chamber}}
Social media provides users with unprecedented access and sharing of information, which has changed how we communicate, discuss, and shape our opinions. The abundance of content allows users to be well informed and open to a broader range of viewpoints. Even though social media gives users more power over the type of content they associate with, there is concern that this could lead to an online filter bubble which then leads to ideological polarisation \cite{57}. Users can aﬀirm their confirmation bias by only following users and consuming content that accords with their previously held beliefs. It automatically filters out the news that a user does not like or disagree with \cite{57}. It can restrict users from accessing various viewpoints and favor creating groups of like-minded users who frame and reinforce a familiar narrative. Thus, leading to the creation of online echo chambers. \\
Echo chambers are environments in which users' opinions, political leanings, or beliefs about a topic are reinforced due to repeated interactions with peers or sources that share similar tendencies and attitudes. For this research, we have adopted the definition
of an echo chamber as put forth by Garimella et al. \cite{5}, i.e. we obtain an echo chamber when ‘the political leaning of the content that users receive from the network agrees with that of the
content they share.' As a result, echo chambers have become a part of social media consumer activity conversations. \\
Filter bubbles can create partisan echo chambers for those who are politically involved and add to an increasing awareness divide between those who are and those who are not. This dynamic is widely seen as harming democracy as a whole and individual citizens' democratic capacities. \\
However, users must allow themselves to venture outside of their comfort zones to access all information and thoroughly analyze their media. Otherwise, users are ingested with much unsubstantiated, misleading information.

\subsection{\textbf{NaMo App}}
\begin{figure}[H]
    \begin{subfigure}[b]{0.49\textwidth}
        \begin{center}
        \includegraphics[width=50mm, height=50mm, frame]{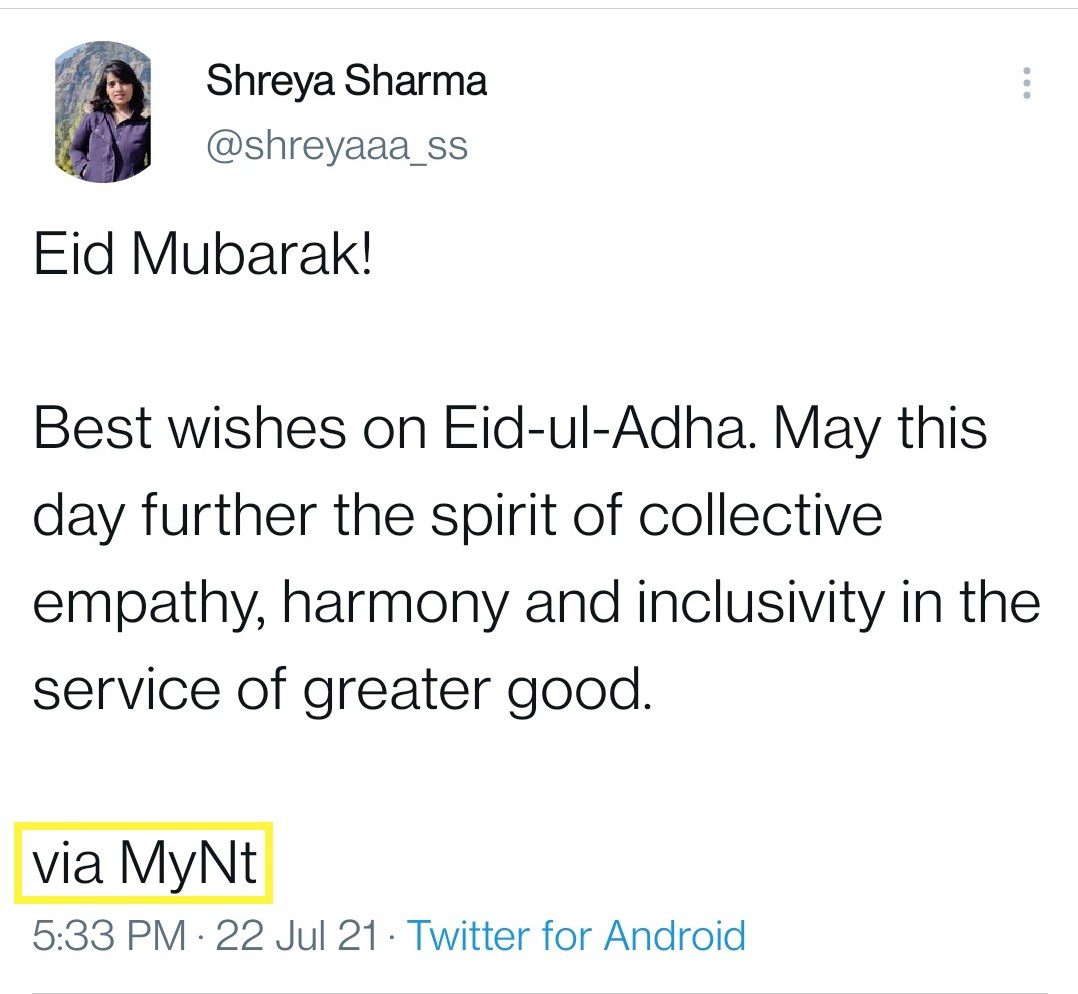}
        \subcaption{Tweet with ‘via MyNt' tag at the end}\label{fig:a}
        \end{center}
    \end{subfigure}
    \begin{subfigure}[b]{0.49\textwidth}
        \begin{center}
        \includegraphics[width=50mm, height=50mm, frame]{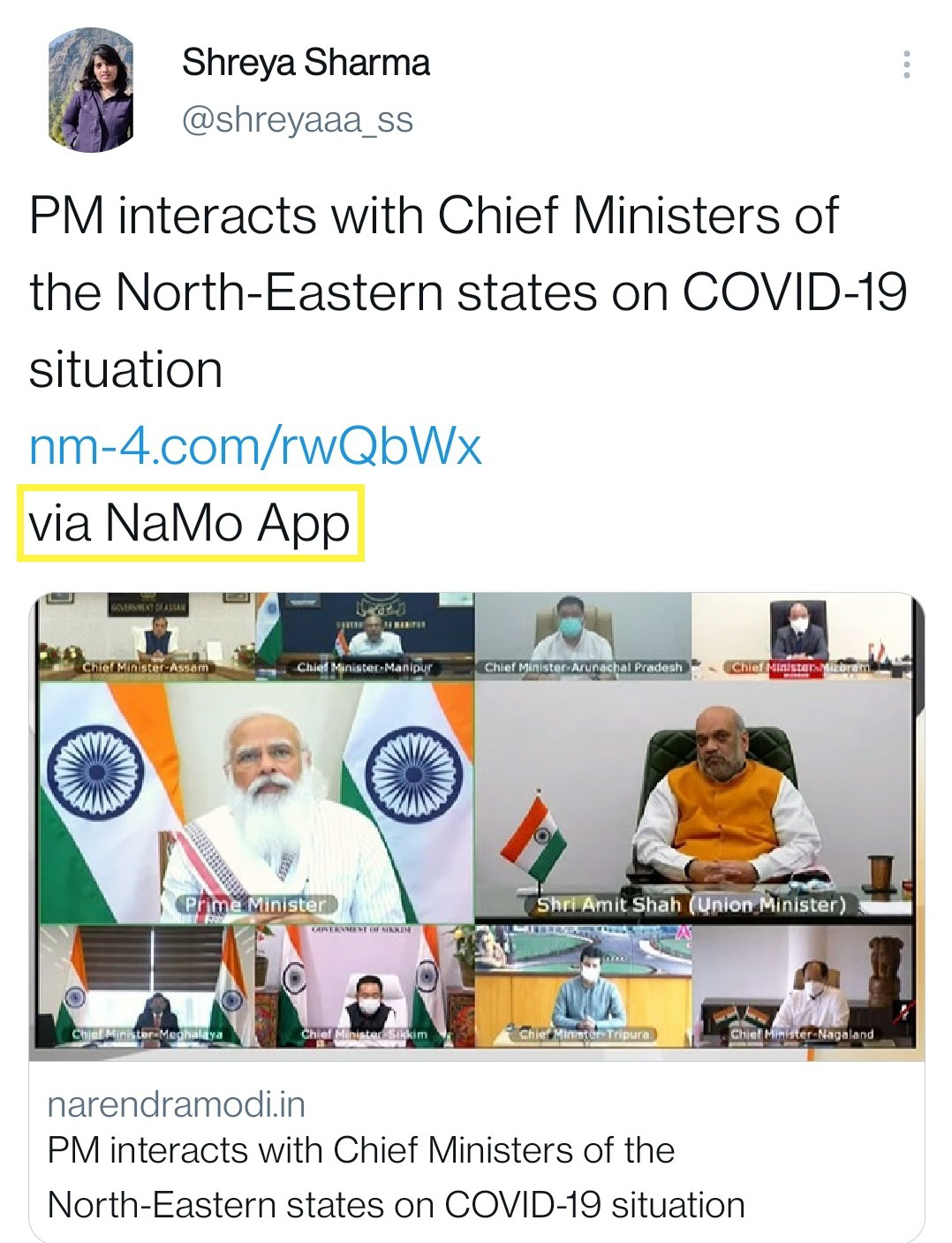}
        \subcaption{Tweet with ‘via NaMo App' tag at the end}\label{fig:b}
        \end{center}
    \end{subfigure}
    \caption{Example of posts that were posted using NaMo App on Twitter.}\label{1}
\end{figure}

In the run-up to the 2014 Lok Sabha polls, Narendra Modi launched his first official app - NaMo App - as part of Modi's Prime Ministerial election campaign \cite{6}. NaMo App is one of the first apps that is centered around a particular political party. The app acted as a digital medium for Bhartiya Janata Party (BJP) for organizing their online political campaign and for the purpose of community outreach. Through NaMo App, users can participate and create party-related tasks expanding from online networking to offline activity initiatives. It also has volunteer activities for which a user gets points and ranks on the leaderboard for every activity they perform on the platform.\\
The app has a  built-in Twitter-like network called ‘My Network' where users can follow other users. Users can also post content in the form of text, images, and videos on the platform. Users can also like, comment, and share posts of other users. It also provides features to share content on other social media platforms like Twitter, Facebook, and WhatsApp. Whenever a user shares content on other socia media platforms, it gets annotated with the ‘via MyNt' or ‘via NaMo App' tag. For instance, whenever any post is shared using NaMo App on Twitter, the tweet gets annotated with ‘with via MyNt' or ‘via NaMo App' tag at its end. (Figure ~\ref{1}). However, users can remove the tags before posting on other social media platforms.\\
According to a report by Bansal \cite{7}, NaMo App provides users with extreme right leaning content and personalized experience based on their location, preferences, and content engagement. Users are shown accounts of BJP leaders and common user accounts of their interest. As this app lacks content moderation, it is open to communal misinformation and fake news.

%% file: Chapters/Chapter2.tex

\section{\textbf{Overview}}

\subsection{\textbf{Problem Statement}}

Our research specifically looks at the users that form online echo chambers on Twitter. We aim to see how an app with extreme right-leaning content affects online conversation on more traditional networks such as Twitter. As Twitter played a significant role in digital engagements between users and political parties during election campaigns, whereas the NaMo app consists of only like-minded users, we aim to see whether the NaMo content receives the same amount of reach and the reaction on Twitter. Also, whether the users that share content using the NaMo App on Twitter are part of online echo chambers.

We analyze the influence of the NaMo App on political environment of Twitter in addition to identifying the presence of online echo chamber on twitter. We aim to address the following research questions:\\
\textbf{RQ1: } How much content on Twitter is contributed by NaMo? What type of content is shared from NaMo App to Twitter ?\\
\textbf{RQ2: } How much influence does NaMo content have on Twitter? \\
\textbf{RQ3: } Can we characterise the users that post NaMo content on Twitter? 
\begin{enumerate}  
\item{Are users part of an Echo Chamber on Twitter?}
\item{Do users of certain states form a larger part of this user group than general users?}
\item{Is there a difference between the users that post using NaMo app and those who post the same content but not using NaMo app?}
\end{enumerate}  
\textbf{RQ4: } What is the reachability of NaMo content on Twitter in comparison to non-NaMo content?

\subsection{\textbf{Datasets}}

To answer the above research questions, we used Twitter datasets of the 2019 Lok Sabha Election, Citizenship Amendment Act (CAA) protests and 2020 Delhi Election. The 2019 Lok Sabha dataset \cite{8} is maintained by the PreCog group at IIIT Delhi, which consists of 45 million tweets collected from January 2019 to May 2019. We also utilized the Twitter dataset of CAA protests consisting of 1.2 million tweets and 2020 Delhi Election consisting of 1.27 million users, collected from December 2019 to February 2020. For the NaMo App, we used a dataset collected and provided by Rohan Rajpal, B.Tech at IIIT Delhi. The data is collected from July 2019 to April 2020. For any other dataset, we used Twitter’s public API.\footnote{\url{
https://developer.twitter.com/en/docs/twitter-api/getting-started/about-twitter-api}}

\subsection{\textbf{Dataset Creation}}

For the 2019 Lok Sabha election and CAA protests dataset, we collected all the tweets posted using the NaMo App, i.e., the tweets annotated with either ‘via MyNt' or ‘via NaMo App' tag at the end. We also collected URLs of all the images that were part of these tweets.

\subsection{\textbf{Definitions}}
\begin{itemize}   
\item \textbf{Seed users:} The users who post content using NaMo App on Twitter, i.e., they
have tweets with ‘via MyNt' or ‘via NaMo App' tag at the end.
\item \textbf{Auxiliary users:} The users who post tweets with content similar to that of
tweets posted via NaMo App but does not use NaMo App for posting them.
\item \textbf{Affected users:} Affected users are a group of seed and auxiliary users.
\item \textbf{NaMo tweets:} Tweets posted using NaMo App on Twitter, i.e., tweets that are annotated either with ‘via MyNt' or ‘via NaMo App' tag at the end.
\item \textbf{Non-NaMo tweets:} Tweets similar to NaMo tweets but not posted using NaMo App on Twitter, i.e., tweets that are not annotated with ‘via MyNt' or ‘via NaMo App' tag at the end.
\end{itemize}

%% file: Chapters/Chapter3.tex

\section{\textbf{Background and Literature Review}} 

Through the advent of social networking platforms, people now have access to a far broader spectrum of viewpoints, vastly expanding the content accessible to citizens and their options over conventional media. The effects of social media sites are studied widely. According to Abbey et al. \cite{44}., social media leads to political polarization and political fragmentation. It can lead to filter bubbles and echo chambers. \\

According to Eli \cite{45}, a filter bubble is a state of the isolated ideological environment that may result from algorithms suggesting personalized content based on the information we agree with, our past behavior, and search history - narrowing down our information spectrum. Echochamber arises as a result of our interaction with communities of people who share similar viewpoints. Echo chambers could be a result of filtering or other processes, but filter bubbles are the result of algorithmic filtering \cite{46}. Individuals are most inclined to seek out evidence that supports their views, but they are less likely to deliberately ignore information that challenges their beliefs \cite{47}. There are studies on partisan websites \cite{48,49}. In India, there have been fewer studies on media being biased \cite{50, 51}. NaMo App is one such app with extremely partisan content \cite{7}.
For our research, we are focusing on echo chambers. Studies on echo chambers and the propagation of partisan content on Twitter are conducted widely. Content and social network structure analysis are used to predict the political alignment of users on social media \cite{9, 10}. Even though, these work claim accuracy rates of over 90 percent, a study by Cohen et al. \cite{11} warns about the drawbacks of such methods and their reliance on politically active users. \\

There have been various studies on polarisation \cite{12,13,14} of Twitter communities over time. Garimella et al. \cite{15} studied how dynamics of polarized topics change under the influence of a sudden increase in user engagement in the topic. Research of Twitter networks \cite{12} has shown that there is segregation that exists in the partisan networks, with the separation becoming quite evident in retweet networks with minimal connectivity between left- and right-leaning users. There are studies to reduce polarisation \cite{53,54,55}. Stasavage et al. \cite{53} explains how public group discussion can help to reduce polarisation but this does not happen when representatives from different factions are involved. Jackson et al. \cite{54} studied that particular users can be removed from the network participation to reduce polarization. A study by Garimella et al. \cite{55} addresses the problem of reducing polarisation by adding edges in the graph of users that produce the most significant reduction in the polarisation. \\

Research done on the existence of echo chambers on Twitter has focused on both network and content analysis. Garimella et al. \cite{5} studied the content consumption and production of twitter users to understand echo chambers, along with how the properties of individual users changes on the basis of their position in the network. Du et al. \cite{13} performed network analysis by comparing network snapshots. They observed that it is more likely that there is addition of intra-community edges and removal of inter-community edges leading to the formation of echo chambers. Another study by An et al. \cite{16} looked at the behaviour of users exposed to political news on Twitter and discovered that 90 percent of the users directly follows news media accounts of only one political leaning, while their followings retweets lead them to diversify their news consumption.\\

However, a study by Morales et al. \cite{17} on Reddit found that cross-cutting political interactions between the two groups are preferred over intra-group interactions, contradicting the echo chamber narrative. Kitchens et al. \cite{52} studied that increased usage of Facebook is linked to greater variety of information sources and a trend toward more polarised news consumption; increased use of Reddit is linked to greater diversity and a move toward more moderate partisan sites; and increased use of Twitter has no significant difference in either. This study came after Twitter changed its algorithm in 2016. \\

There have also been various studies \cite{20,21,22,23,24,25} to understand how one social media platform can influence the content on other social media platforms using Hawkes process \cite{26}. Zannettou et al. \cite{22} studied how different web communities such as Twitter, Reddit, and 4chan are influenced by each other in the context of posting mainstream and alternate news sources. Another study \cite{20} looked at the influence Twitter, Reddit, 4chan, and Gab have on each other in the spread of memes. A study by Zannettou et al. \cite{21} studies quantify state-sponsored accounts' influence on popular Web communities like Twitter, Reddit, 4chan's Politically Incorrect board (/pol/), and Gab, with respect to the dissemination of images.

%% file: Chapters/Chapter4.tex

\section{\textbf{Preliminary Experiments}} 

To gain a better understanding of the 2019 Lok Sabha Election and CAA dataset, we did a basic content analysis in the form of a word cloud as well as some other visualisation.
\subsection {Word Cloud Analysis}
\begin{figure}[h]
    \begin{subfigure}[b]{0.49\textwidth}
        \begin{center}
        \includegraphics[width=50mm, height=50mm, frame]{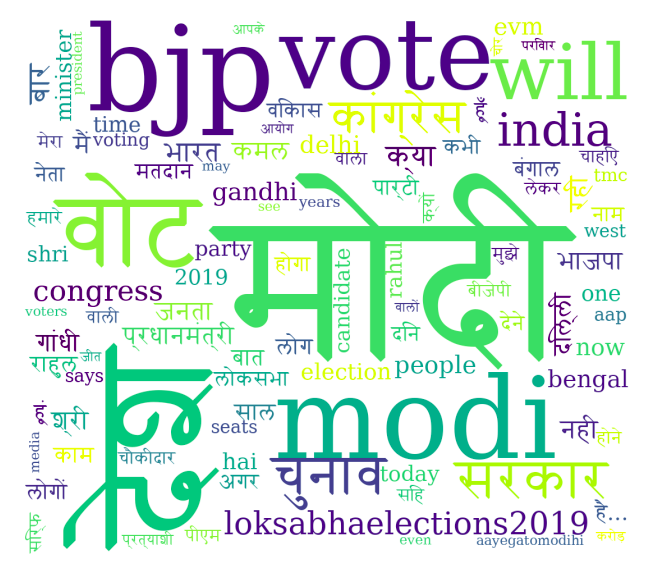}
        \subcaption{For 2019 Lok Sabha Election}\label{fig:2}
        \end{center}
    \end{subfigure}
    \begin{subfigure}[b]{0.49\textwidth}
        \begin{center}
        \includegraphics[width=50mm, height=50mm, frame]{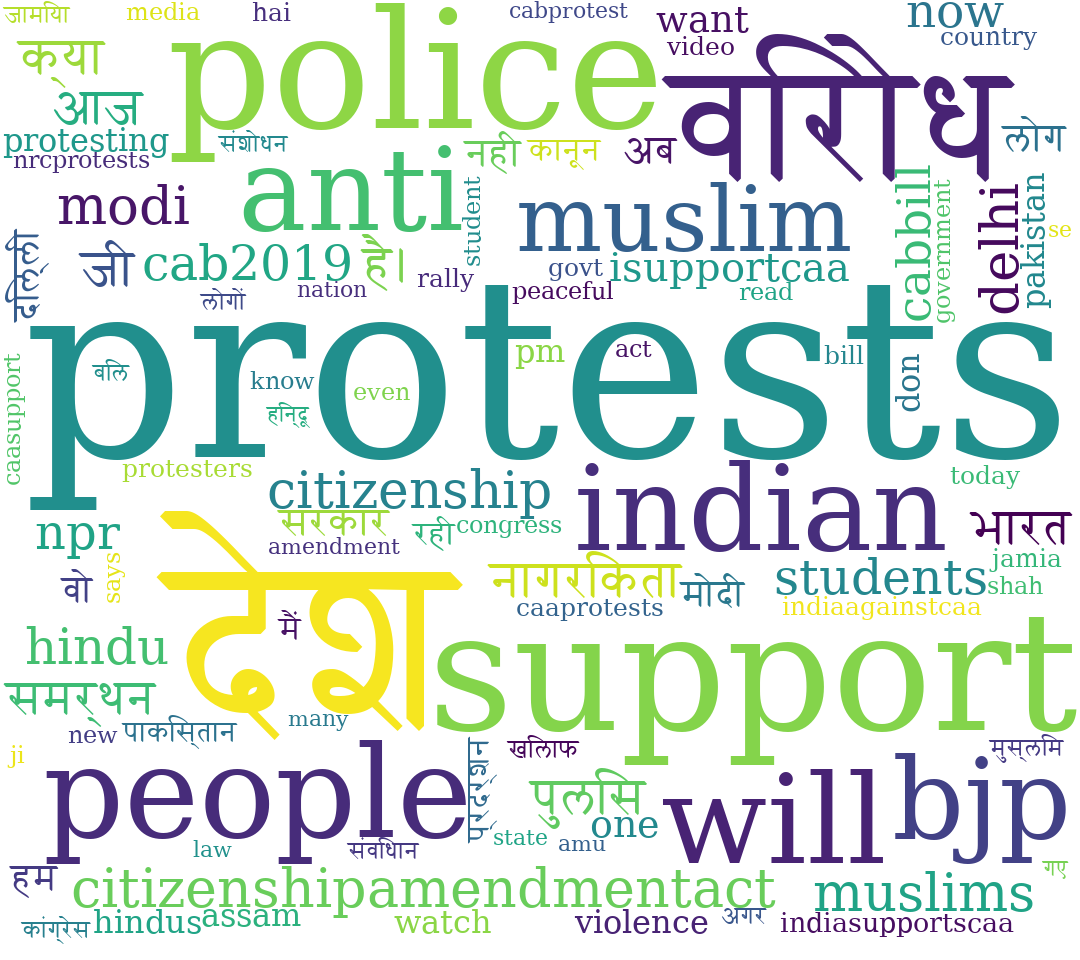}
        \subcaption{For CAA protests}\label{fig:3}
        \end{center}
    \end{subfigure}
    \caption{Wordcloud showing the most frequent words in our datasets.}\label{2}
\end{figure}

To get a broader overview of what topics are more discussed, we created a word clouds. The word cloud is a representation of the most frequent words in the dataset. In a word cloud, the larger and bolder the word is, the more often it appears in the data and the more significant it is than the words with less occurance. This frequency-based word cloud method is beneficial for visually summarising large volumes of data. However, word clouds do not take into account the context in which the word occurs.\\
In the obtained word cloud (Figure ~\ref{2}), we observed that BJP oriented terms such as ‘BJP', ‘modi' are more frequent than words related to other political parties such as ‘congress', ‘gandhi', etc, indicating the BJP wave during the 2019 Lok Sabha election. Other neutral election-related terms also include ‘vote', ‘loksabhalections2019', ‘desh', etc. In Figure ~\ref{fig:3} for CAA protests, we removed common words like CAA, CAB, NRC, and India to get more clear idea. We observed words like ‘protests', ‘support', ‘anti', ‘muslim’, ‘{\dn EvroD}’, etc.
The frequency of ‘support' and ‘{\dn EvroD}' are almost equal.

\subsection{\textbf{Hashtag Analysis}}
\begin{figure}[H]
    \begin{subfigure}[b]{1\textwidth}
        \begin{center}
        \includegraphics[width=130mm]{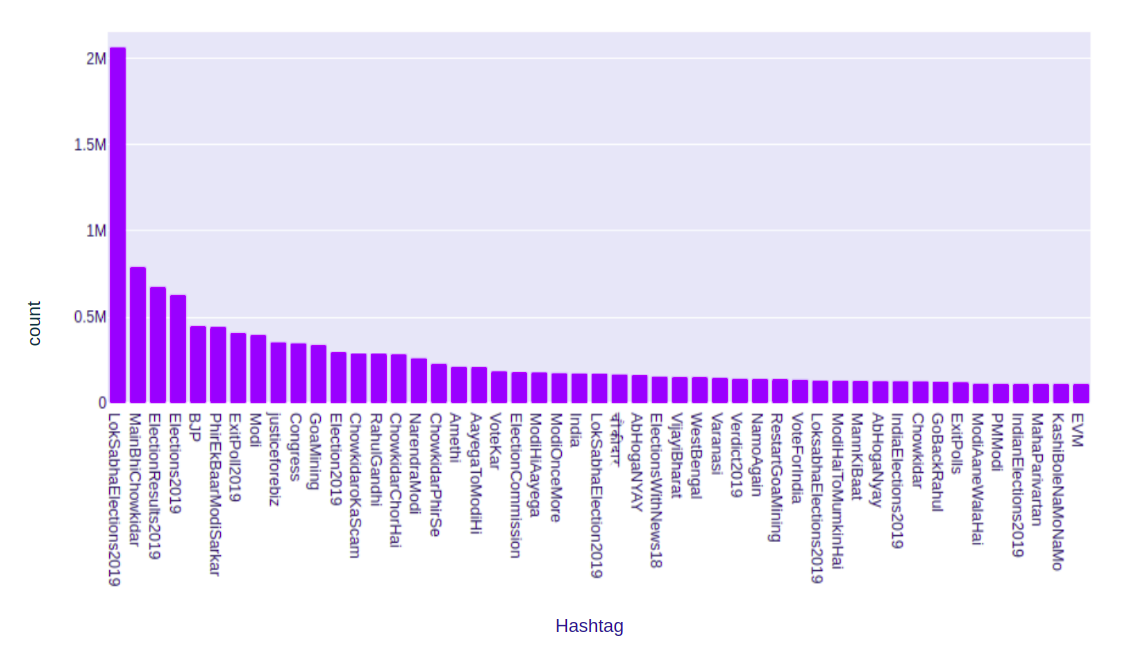}
        \subcaption{For 2019 Lok Sabha Election}\label{fig:electiona}
        \end{center}
    \end{subfigure}
    \begin{subfigure}[b]{1\textwidth}
        \begin{center}
        \includegraphics[width=130mm]{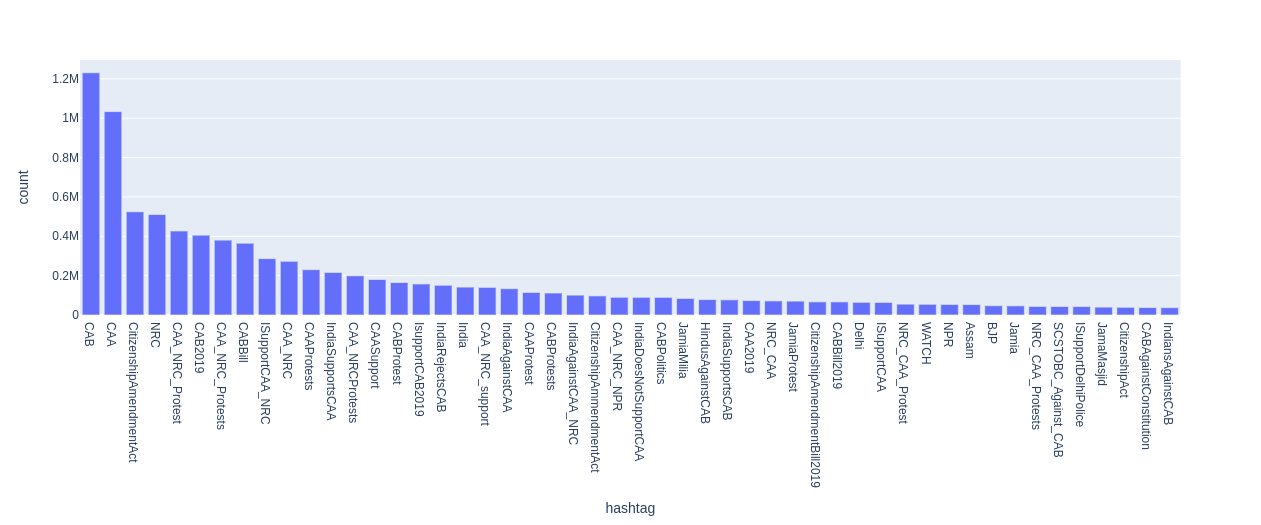}
        \subcaption{For CAA protests}\label{fig:caab}
        \end{center}
    \end{subfigure}
    \caption{Most frequent hashtags}\label{2hashtag}
\end{figure}

Hashtags are used in social media to help users interested in a particular topic to find it by searching for a keyword. It helps in attracting attention to a user's posts and encourages engagement. We created a plot that shows the most frequent hashtags used during the 2019 Lok Sabha elections and during CAA protests. Apart from neutral hashtags, we observed that hashtags related to the BJP party and the chowkidaar movement are the most tweeted hashtags indicating the BJP wave during 2019 Lok Sabha Election(Figure ~\ref{fig:electiona}). For CAA protests (Figure ~\ref{fig:caab}), we observed hashtags which supports CAA have more frequency than hashtags which are anti-CAA.

\subsection{\textbf{User Metadata Analysis}}
According to studies \cite{28}, a user's political leaning on social media sites can be determined by their profile change behavior. Twitter users often change their screen\_names or description\footnote{\url{https://developer.twitter.com/en/docs/twitter-api/v1/data-dictionary/overview/user-object}} to support a political campaign or a political leader.  For instance, during the Chowkidaar campaign, many
users who supported BJP changed their screen\_names by adding Chowkidar to it \cite{29}. To analyze such trends, we created a plot that shows the number of users that changed their screen\_name on a given date. The peak observed on 18 April 2019 corresponds to the second phase of elections polling \cite{27}. Similarly, we also observed a peak on 23 May 2019 - the day election results were announced \cite{27,2}. Moreover, we also observed that the change in screen\_name is not limited to BJP but is also seen among users affiliated with other political parties.

\begin{figure}[H]  
\centering
\includegraphics[angle=270,width=1\textwidth]{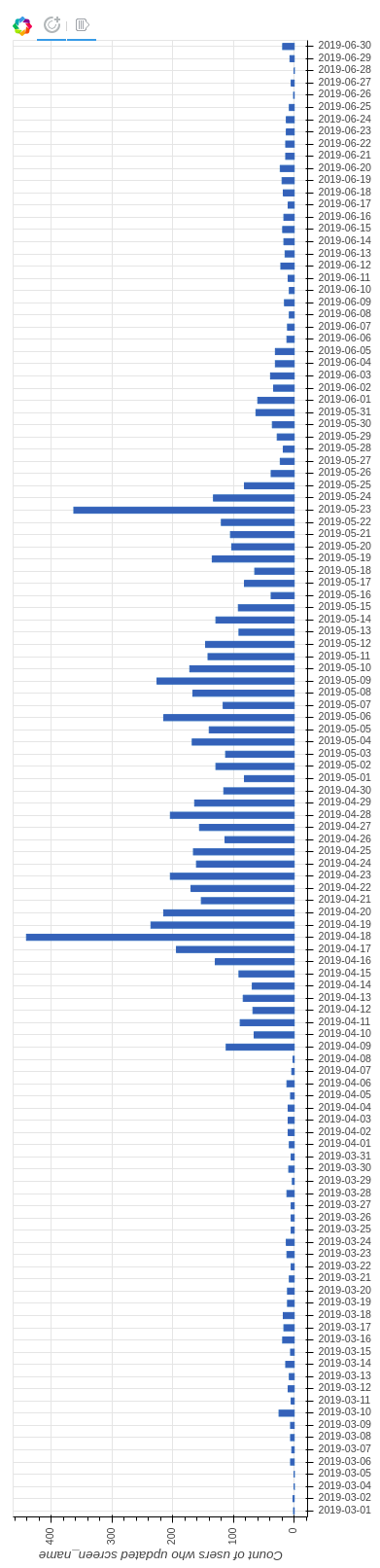}
\caption{User's metadata activity during the 2019 Lok Sabha Elections.}
\end{figure}

%% file: Chapters/Chapter5.tex
\section{\textbf{Echo Chambers during 2019 Lok Sabha Elections}} \label{Echo Chamber during Indian Elections}

The way people engage with one another and consume information has changed as a result of social media. As a facilitator for public opinion and debate, social media has become increasingly important, creating an online echo chamber. Understanding online polarisation is critical because it can negatively impact politics, democratic decision-making, and societal life in general. Users receive biased information as a result of polarisation. According to Stewart et al. \cite{35}, echo chambers also affect users' voting habits.\\ 
In this chapter, we analyze the presence of echo chambers during the 2019 Lok Sabha elections on Twitter by performing content and network analysis. In content analysis, we partitioned the users based on their political affiliation by analyzing the polarities of the content users produce and consume. In network analysis, we analyzed the user's followers, following, and retweet network.\\
Since we are specifically focusing on users who form echo chambers and are active on Twitter, we filtered out those users who have tweeted at least 50 tweets during the 2019 Lok Sabha Election and 2020 Delhi Election and have used the most frequent hashtags. After filtration, 6037 common users were found from the list of users that have tweeted more than 50 times and had participated in the popular hashtags from these datasets.

\subsection{\textbf{Partitioning of Users}}
We followed the below steps for partitioning the users according to their political affiliations:
\begin{enumerate}   
\item{\textbf{NivaDuck Dataset:}}\\
We used the NivaDuck dataset \cite{31} to assign affiliations for the users. It consists of Twitter handles of various Indian and US political leaders along with their party labels. We had 243 users in common with this dataset.
\item{\textbf{User Metadata:}}\\
As user’s political leaning on social media sites can be determined by their profile change behavior \cite{28}, we utilised a user's screen\_name and twitter descriptions\footnote{\url{
https://developer.twitter.com/en/docs/twitter-api/v1/data-dictionary/overview/user-object}} to denote their political inclinations. In this step, we were able to find 474 users. We also did a manual verification of their twitter accounts to ensure correct affiliation annotations. We used the following keywords:
\begin{itemize}
\item{\textbf{BJP:}}\\
– Keywords in screen\_name: bjp, modi, namo\\
– Keywords in description: bjp, modi, namo, narendra modi, bhartiya janta
party, amit shah, amitshah, narendramodi, bjp4india
\item{\textbf{INC:}}\\
– Keywords in screen\_name: congress, inc, cong\\
– Keywords in description: rahul gandhi, congress, inc,  priyanka gandhi, raga, shashi tharoor
\end{itemize}

\begin{table}[!htb]
  
  \label{table:1}
  \begin{tabular}{ccl}
    \toprule
    \textbf{screen\_name} & \textbf{political affiliation} \\
    \midrule
   
    MODIfied\_SKP & BJP \\
\hline
Shivam\_INC & Congress \\
\hline
Krish\_BJP & BJP \\
\hline
amitsoni\_INC & Congress \\

  \bottomrule
  
\end{tabular}
\caption{Examples of the users assigned a political affiliations according to their screen\_name.}
\end{table}

\begin{table}[!h]
  
  \label{table:2}
  \begin{tabular}{p{3cm}|p{5cm}|p{3cm}}
    \toprule
    \textbf{screen\_name} & \textbf{description} & \textbf{political affiliation}\\
    \midrule
    
 Dilsedesh & Lawyer \& Not associated with Congress in any manner My tweets are my personal views and still Rahul Gandhi is my leader
 & Congress \\
 \hline
 ParasKGhelani & RSS, ABVP, BJP Research Analyst Stock Market Corporate Finacial Analyst Finance, Sports. Cyber Security Data Protection Tweets \& RTs are personal. & BJP \\
 \hline
 sparjaga & Hardcore MODIJI fan. Former RSS pracharak. District secretary BJP IT \& Social media Tirupattur DT.Tamil nadu. & BJP \\
 \hline
 Nil\_deshbhratar & President Central Nagpur Congress Social Media Cell @INCIndia @INCMaharashtra & Congress \\
 
  \bottomrule
\end{tabular}
\caption{Examples of the users assigned political affiliation according to their description.}
\end{table}

\item{\textbf{Hashtag Usage of the User}}\\
For partitioning of the remaining 5320 twitter users, we utilised their hashtag usage as proxy
of their political leanings. We began with annotating 600 political hashtags\footnote{\href{https://docs.google.com/spreadsheets/d/1l6yRFDGkY9jmUWNCgDdKEvnp55VkeMA_HyoYkWs33xg/edit?usp=sharing}{2019 Lok Sabha Election Dataset Annotated Hashtags}}
that trended during the 2019 Lok Sabha Elections. We annotated them as pro-bjp, anti-bjp, pro-congress,
anti-congress and neutral.
Then we calculated the following attributes for each user:
\begin{enumerate}   
\item{\textbf{pro bjp ratio:}} non-unique count of pro-bjp hashtags/number of annotated hashtags
\item{\textbf{pro congress ratio:}} non-unique count of pro-congress hashtags/number of annotated hashtags
\item{\textbf{anti bjp ratio:}} non-unique count of anti-bjp hashtags/number of annotated hashtags
\item{\textbf{anti congress ratio:}} non-unique count of anti-congress hashtags/number of annotated hashtags
\item{\textbf{percent of hashtags used:}} number of annotated hashtags/not neutral
\item{\textbf{number of annotated hashtags:}} non-unique count of pro-bjp, pro-congress, antibjp, anti-congress hashtags
\item{\textbf{not neutral:}} non-unique count of hashtags that are not neutral (i.e. pro-bjp, procongress, anti-bjp, anti-congress and no label)
\item{\textbf{total hashtags:}} non-unique count of hashtags used by a user
\end{enumerate} 
We have set a minimal threshold for percentage of hashtags utilized to ensure that we don't assign political affiliation to users who don't have enough annotated hashtags. After selecting the users who met this criterion (0.1), we were left with 3920 users. The users were then allocated a leaning based on the ratio with the highest value.\\

\begin{figure}[H]  
\centering
\includegraphics[width=140mm]{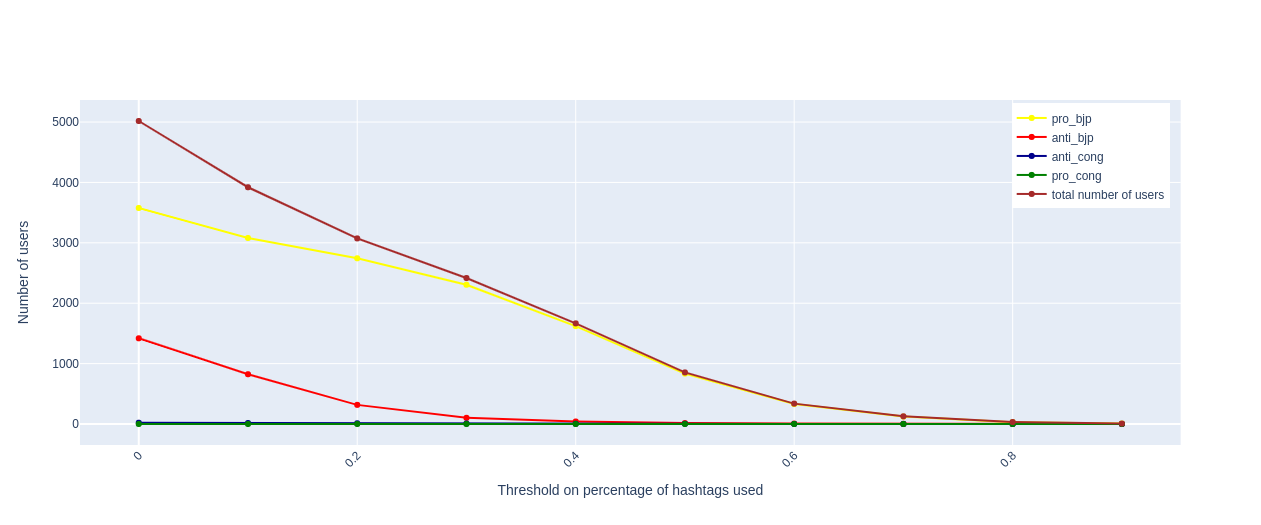}
\caption{Partitioning of users into different political affiliations according to change in the threshold
on percent of hashtags used.}
\end{figure}
\end{enumerate}

\subsection{\textbf{Content Analysis}}
To identify the presence of echo chambers, we used the similar methodology as outlined by Garimella et al. \cite{5}:

\subsubsection{Methodology}
\begin{figure}[H]
    \begin{subfigure}[b]{0.49\textwidth}
        \begin{center}
        \includegraphics[width=60mm, height=50mm]{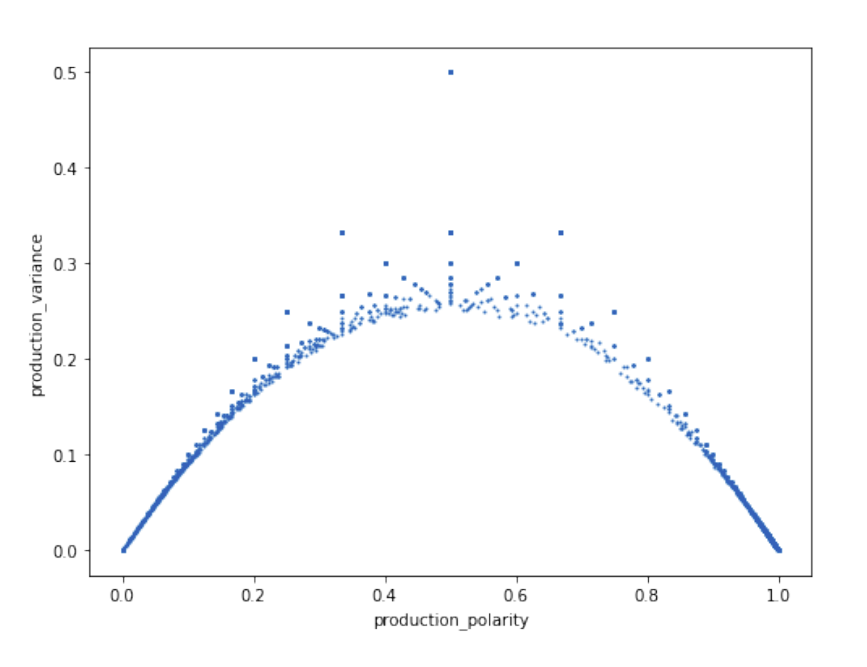}
        \subcaption{Production Variance vs Production Polarity}\label{3}
        \end{center}
    \end{subfigure}
    \begin{subfigure}[b]{0.49\textwidth}
        \begin{center}
        \includegraphics[width=60mm, height=50mm]{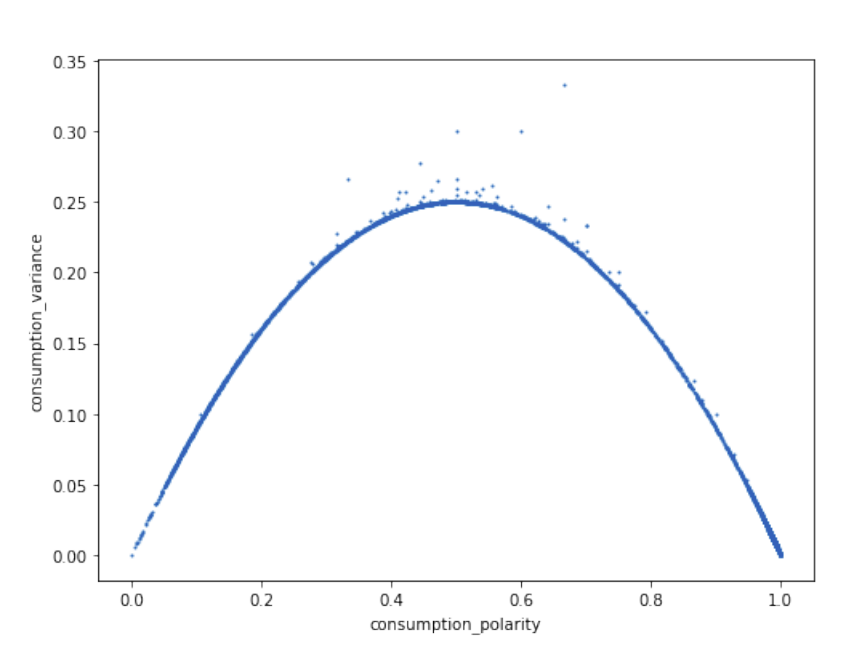}
        \subcaption{Consumption Variance vs Consumption Polarity}\label{fig:4}
        \end{center}
    \end{subfigure}
    \caption{Graph for Variance vs Polarity.}\label{4}
\end{figure}

\begin{figure}[H]
    \begin{subfigure}[b]{0.49\textwidth}
        \begin{center}
        \includegraphics[width=60mm, height=50mm]{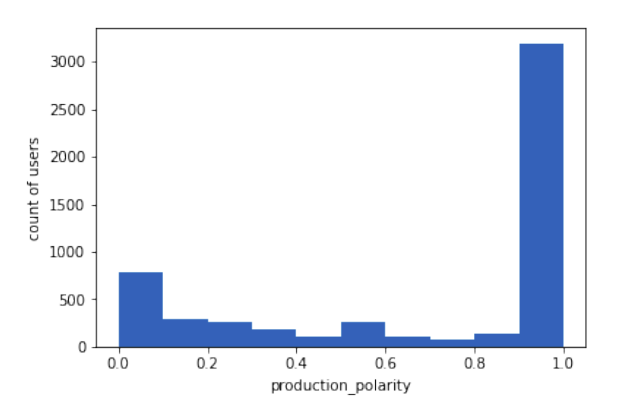}
        \subcaption{Distribution of Production Polarity}\label{fig:a}
        \end{center}
    \end{subfigure}
    \begin{subfigure}[b]{0.49\textwidth}
        \begin{center}
        \includegraphics[width=60mm, height=50mm]{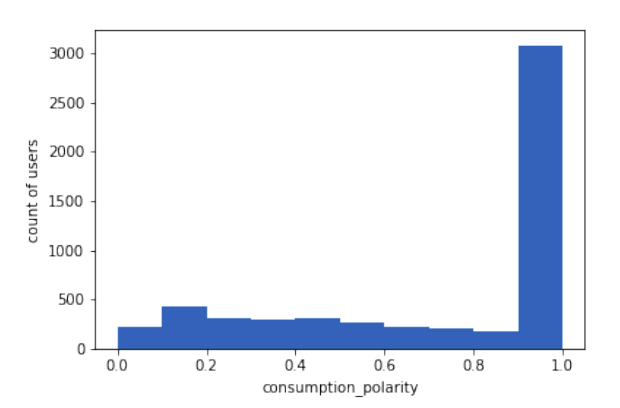}
        \subcaption{Distribution of Consumption Polarity}\label{fig:b}
        \end{center}
    \end{subfigure}
    \caption{Distribution of polarities of users.}\label{distri}
\end{figure}

We assigned a score of 0 or 1 to each hashtag for the values listed below, such that an anti-BJP hashtag or a pro-Congress hashtag gets a value 0 and a pro-BJP hashtag gets a score of 1. We calculated the following values for each user in our database:
\begin{enumerate}   
\item{\textbf{Production polarity:}} The average of the
scores of all the hashtags that are utilised by that user.
\item{\textbf{Production variance:}} The variance of
the scores of all the hashtags that are utilised by that user.
\item{\textbf{Consumption polarity:}} The average
of the scores of all the hashtags that are utilised by all the users that user u follows.
\item{\textbf{Consumption variance:}} The variance
of the scores of all the hashtags that are utilised by all the users that user u follows.
\end{enumerate}  
Then, we plotted the graphs for variance vs. polarity (Figure ~\ref{4}) and distribution of production and consumption polarities of users (Figure ~\ref{distri}).

\subsubsection {Observations}
\begin{figure}[H]  
\centering
\includegraphics[width=120mm]{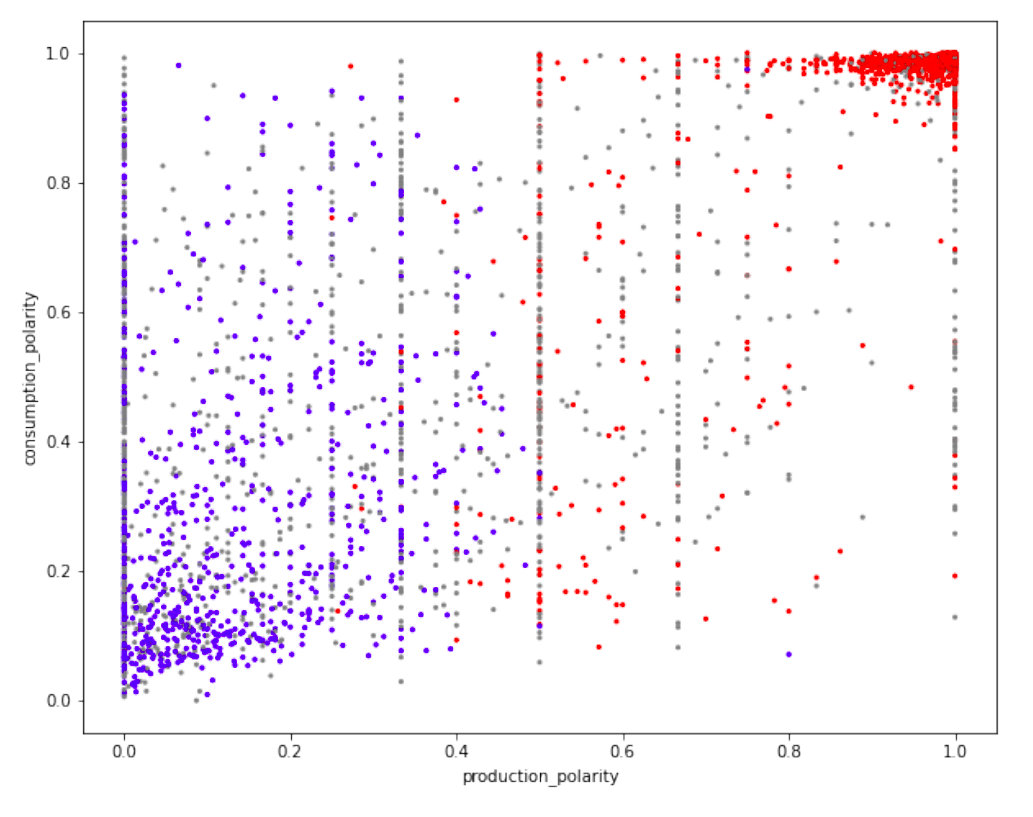}
\caption{Consumption Polarity vs Production Polarity} \label{5}
\end{figure}

For each user in Figure ~\ref{3}, we investigated the link between production polarity and production variance. As per our findings, production variance tends to be near to zero for users at the extreme ends of production polarity. Users with production polarity close to 0.5, on the other hand, have the highest production variance. As a result, a downward-U shape emerges. This demonstrates that users who create content with an average polarity close to 0 or 1, have a very little variation in the polarity of the content they create. When comparing consumption polarity and consumption variance, we see a similar pattern in Figure ~\ref{fig:4}. Therefore, users who consume information with an average polarity close to 0 or 1 have very little variation in the polarity of the tweets they consume.\\
In addition, we also plotted the distribution of degrees in Figure ~\ref{distri}. We observed that the number of users with polarity close to 1 are more than the number of users with polarity tending to 0. Because we assigned a score of 1 to the content that a user produces or consumes, we can infer that the filtered users are more BJP-centric.\\
In Figure ~\ref{5}, we plotted the values of production polarity and consumption polarity for each user to see the relationship between two properties. The BJP and Other users are colored red and blue, respectively. We infered that users who favor the BJP party tend to cluster in the top-right corner of the plot, whereas ‘Other Users' cluster in the bottom-left corner. Considering pro-Congress and anti-BJP hashtags have a value of 0 and pro-BJP hashtags have a value of 1, figure ~\ref{5} indicates that people with a particular political affiliation tend to produce and consume tweets that reflect their ideology/beliefs.

\subsection{\textbf{Network Analysis}}
We also performed network analysis to see the presence of different communities on Twitter. 
\subsubsection{Definitions}
To visualise different network formed among the users in our database, we used the network analysis tool Gephi\footnote{\url{
https://gephi.org/}}.

\begin{itemize}
\item{\textbf{Follow Network:}} This is a directed network such that, if a user \textit{u} follows a user \textit{v}, we
have an edge from \textit{v} to \textit{u}.
\item{\textbf{Friends Network:}} This is a directed network such that, if a user \textit{u} is followed by a user \textit{v}, we
have an edge from \textit{v} to \textit{u}.
\item{\textbf{Retweet Network:}} This is a directed network such that, if a user \textit{u} retweets a user \textit{v},
we have an edge from \textit{v} to \textit{u}.
\end{itemize}

\subsubsection{Methodology}
We performed the below steps to create the visualisations for analysing the Network:
\begin{enumerate}   
\item{We utilized the Force Atlas 2 layout in Gephi to better visualize the graph after importing the graph dataset. It performs:\\
 - Scaling: Control scale of the expansion of the graph.\\
 - Dissuade hubs: Applies stronger repulsive forces to opposing hubs.\\
 - Prevent overlap: Prevent nodes from overlapping.}
\item{We calculated the graph's Modularity, which partition the network into
communities of densely connected nodes, with the nodes belonging to different
communities being only sparsely connected.}
\item{The graph's nodes are colored on the basis of political affiliation of users. We looked at three types of users: BJP supporters, Other Users (anti-BJP and pro-Congress), and Unknown. The purple, green, and orange represent the BJP, Other, and Unknown nodes, respectively in figure ~\ref{6}, figure ~\ref{friends} and in figure ~\ref{7}. }
\item{We scaled the size of the nodes according to their degree.}
\item{We utilized the Circle Pack layout with modularity and degree as the parameters to rearrange the nodes of the graph so that nodes with the same modularity class are closer to one another.}
\end{enumerate}

\subsubsection{Observations}

\begin{figure}[h]  
\centering
\includegraphics[width=1\linewidth, frame]{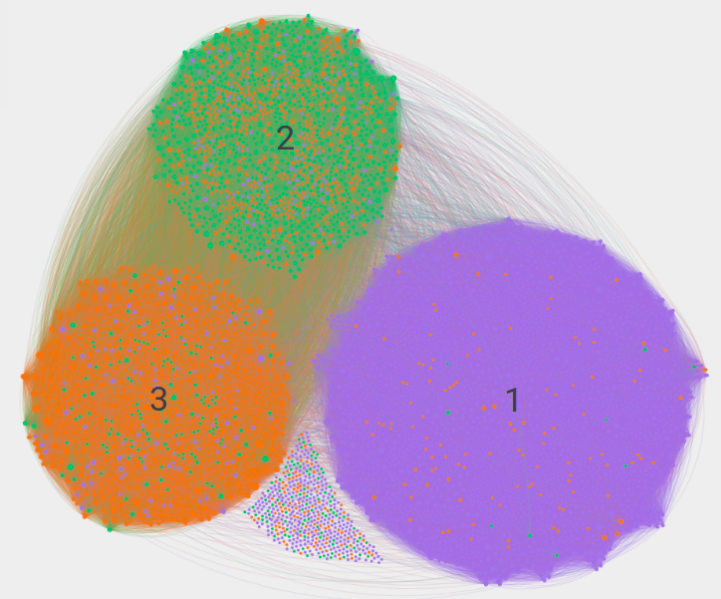}
\caption{Graph of the follow network graph that we created using Gephi. The nodes belonging
to BJP, Other and Unknown are of the color purple, green and orange respectively.}\label{6}
\end{figure}

In follow network (Figure ~\ref{6}), we observed that the communities created using modularity have a significant overlap with the political affiliations of the user that we had assigned. We also observed three main communities generated by modularity in the graph for the follow network. There are also some orange points, i.e. users whose affiliations are unknown to us in this community. The majority of users with affiliation Other, i.e., users who are pro-Congress and anti-BJP, are found in community 2. The formation of separate communities of left-leaning (anti-BJP and pro-Congress) and right-leaning (pro-BJP) users demonstrates these varied political leanings are transferred into social media networks in the form of tightly-knit communities.

\begin{table}[h]
    \centering
\begin{tabular}{ p{3cm}|p{3cm}|p{3cm}|p{3cm} }
 \toprule
 \textbf{Community} & \textbf{Pro-BJP} & \textbf{Other} & \textbf{Unknown}\\
 \midrule
 
 Community 1 & 0.958 & 0.003 & 0.039\\
 \hline
 Community 2 & 0.088 & 0.534 & 0.378\\
 \hline
 Community 3 & 0.098 & 0.128 & 0.774\\
 
 \bottomrule
\end{tabular}\\
\caption{Fraction of users belonging to a given affiliation in a given community in Follow Graph.}
\label{table:3}
\end{table}

\begin{figure}[h]  
\centering
\includegraphics[width=1\linewidth, frame]{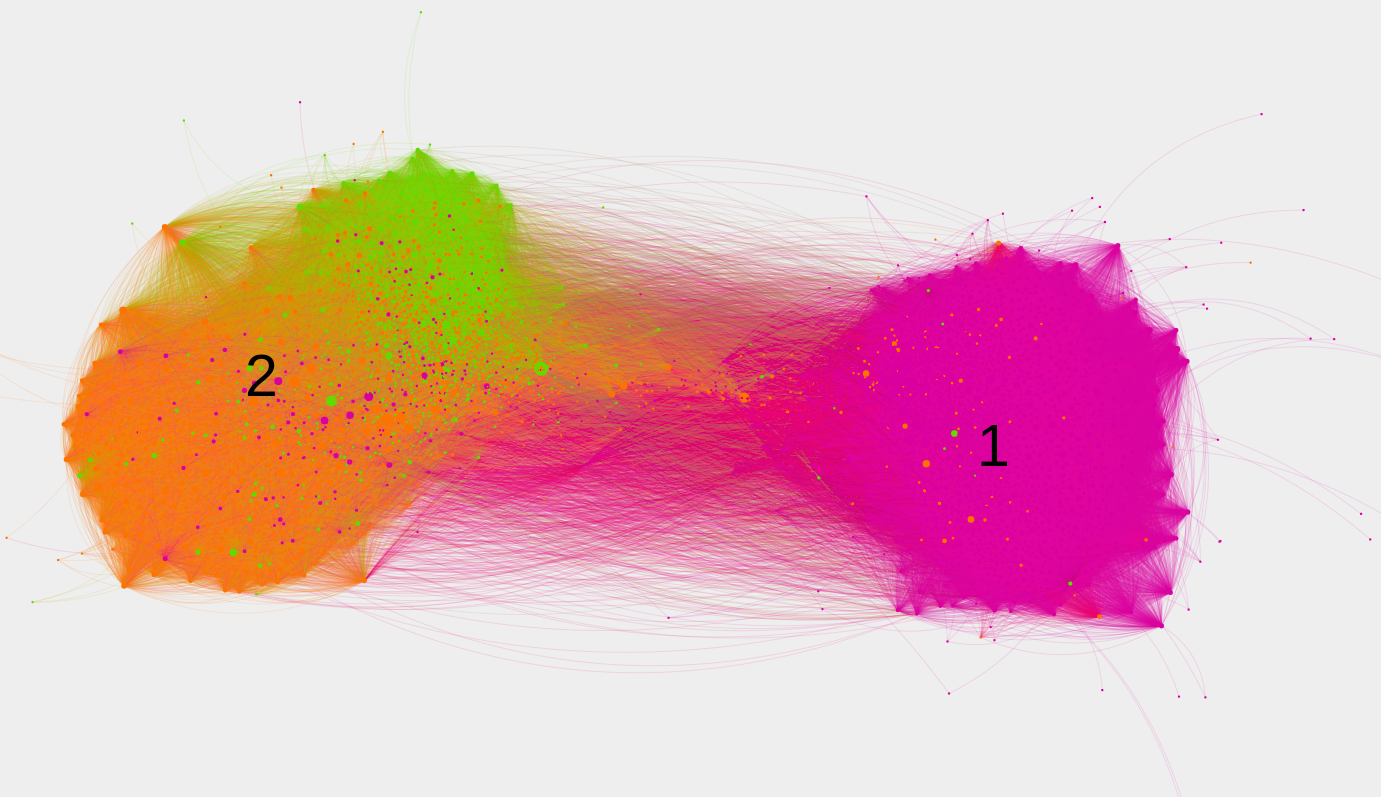}
\caption{ Graph of the friends network graph that we created using Gephi. The nodes belonging
to BJP, Other and Unknown are of the color purple, green and orange respectively.} \label{friends}
\end{figure}

In Figure ~\ref{friends} of friends network, we observed 2 communities generated using modularity. Community 1 is dominated by Pro-BJP users, whereas Community 2 comprised more of Other users. The graph can be used as an indicator that users are more followed by users who matches with their own political affiliations - have a more closer connection - and thus, forming a greater community. The results are identical to the ones we observed in follow network.

\begin{table}[h]
    \centering
\begin{tabular}{ p{3cm}|p{3cm}|p{3cm}|p{3cm} }
 \toprule
 \textbf{Community} & \textbf{Pro-BJP} & \textbf{Other} & \textbf{Unknown}\\
 \midrule
 
 Community 1 & 0.957 & 0.003 & 0.040\\
 \hline
 Community 2 & 0.08 & 0.306 & 0.614\\
 
 \bottomrule
\end{tabular}\\
\caption{Fraction of users belonging to a given affiliation in a given community in Friends Graph.}
\label{table:3}
\end{table}

\begin{figure}[h]  
\centering
\includegraphics[width=1\linewidth, frame]{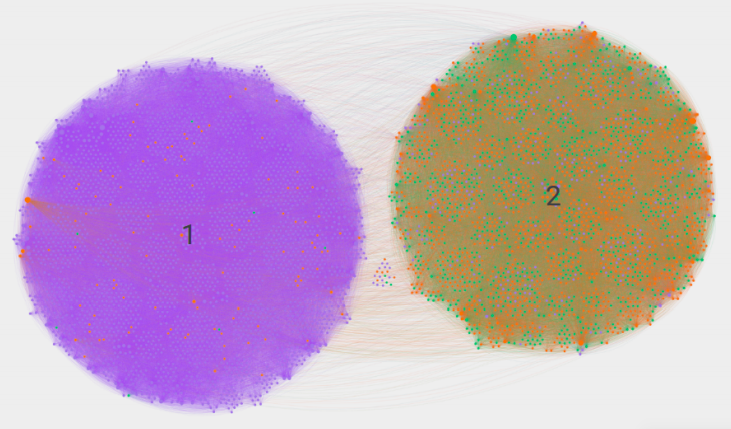}
\caption{ Graph of the retweet network graph that we created using Gephi. The nodes belonging
to BJP, Other and Unknown are of the color purple, green and orange respectively.} \label{7}
\end{figure}

In Figure ~\ref{7}, we observe that in graph for retweet network, there are 2 main communities that are generated using modularity. The retweet network graph, like the follow network graph, displays a high overlap between users from community 1 and those from the BJP. The other community is made up of users with Other affiliations (anti-BJP and pro-Congress) as well as users whose affiliations are unknown. This indicates that these two groups of users in the retweet network have a greater connection.

\begin{table}[h]
\centering
\begin{tabular}{ p{3cm}|p{3cm}|p{3cm}|p{3cm} }
\toprule
 \textbf{Community} & \textbf{Pro-BJP} & \textbf{Other} & \textbf{Unknown}\\
 \midrule

  Community 1 & 0.964 & 0.002 & 0.034\\
  \hline
  Community 2 & 0.096 & 0.368 & 0.536\\

\bottomrule
\end{tabular}\\
\caption{Fraction of users belonging to a given affiliation in a given community in Retweets Graph.}
\label{table:3}
\end{table}

%% file: Chapters/Chapter6.tex

\section{Contribution of the NaMo App to
Twitter content} \label{Contribution of the NaMo App to
Twitter content} 


To answer our RQ1 - How much content on Twitter is contributed by NaMo? What type of content is shared from NaMo App to Twitter? - We collected all the tweets posted using NaMo App. To gather tweets that were similar to the NaMo tweets, we have used image and text clustering.

\subsection{Text Clustering}

\subsubsection{Preprocessing}
We applied the following pre-processing techniques on the tweets:

\begin{enumerate} 

\item{The text was converted into lowercase.}
\item{All punctuation, twitter handles, emojis and links were removed from the text.}
\item{All stopwords were removed.}
\end{enumerate}

\subsubsection{Clustering}
\begin{enumerate}   
\item{We used CountVectorizer\footnote{\url{
https://scikit-learn.org/stable/modules/generated/sklearn.feature\_extraction.text.CountVectorizer.html}}
to create vectors for all the tweets such that any vocabulary
word with document frequency 1 was removed}
\item{We then removed all tweets with length less than 5.}
\item{We performed K-Means clustering\footnote{\url{
https://scikit-learn.org/stable/modules/generated/sklearn.cluster.KMeans.html}}
such that the model was trained on the normalized
vectors of the tweets that have ‘via MyNt' or ‘via NaMo App' at the end.}
\item{For the tweets remaining after step 2, we only considered a match if euclidean distance
between the normalized vector of the tweet and its cluster centroid was less than 0.45}
\end{enumerate} 

We obtained 4170 tweets in the Lok Sabha Election 2019 dataset that matched the tweets that
were made using NaMo App. For the CAA Protest dataset, we were able to find 20500 such
tweets.

\subsection {Image Clustering}
\begin{figure}[h]  
\centering
\includegraphics[width=1\linewidth, frame]{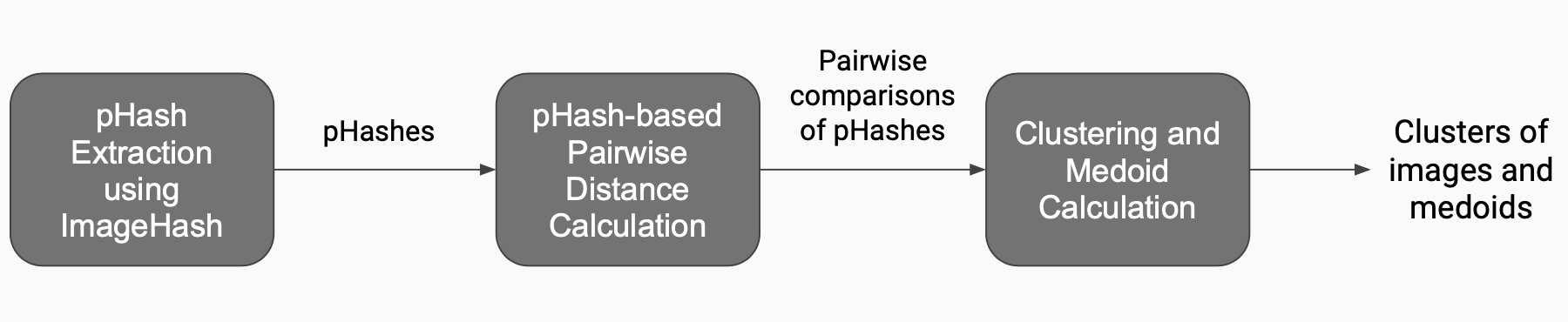}
\caption{Methodology for image clustering}\label{methodo}
\end{figure}

To get the visually similar images, we used the methodology (Figure ~\ref{methodo}) described in paper \cite{20}.

\begin{enumerate}   
\item{We calculated the Perceptual Hashing (pHash) algorithm \cite{33} for each image using the ImageHash library\footnote{\url{https://pypi.org/project/ImageHash/}} present in the tweets that posted using the NaMo App. Through this hashing, any images similar to the human eye map to a similar hash value. Similar images have minor differences in their vector.}
\item{We calculated the pairwise hamming distance between the hash of each image to the hash of every other image.}
\item{We clustered the images having a distance less than 10 between them, using the DBSCAN algorithm because it performs well over noisy data.}
\item{For each cluster, we also calculated the medoid, which is the image that has the minimum average Hamming distance from all the images in the cluster.}
\item{We then calculated the Perceptual Hash for every image that was not present in tweets posted using NaMo App. We assigned them to the cluster with minimum Hamming distance from the cluster medoid.}
\end{enumerate} 
For Lok Sabha Election 2019 dataset, We were able to find 4705 distinct images that similar to the images present in the tweets posted with the NaMo App.

\begin{figure}[h]  
\centering
\includegraphics[width=1\linewidth, frame]{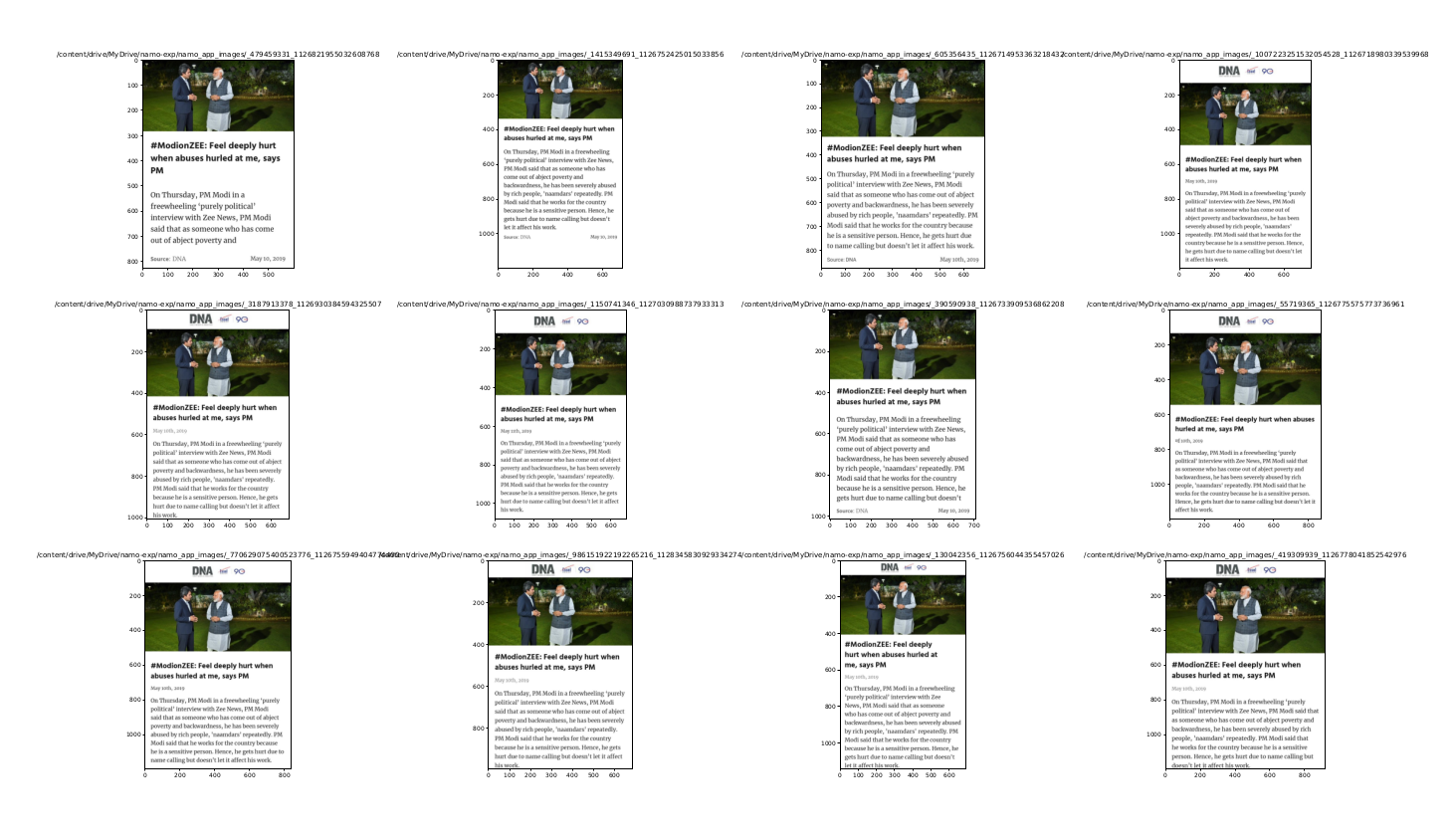}
\caption{Example of a cluster}
\end{figure}

\subsection{Temporal Analysis}
We also performed temporal analysis to see if tweets similar to the content posted from the NaMo App were posted first on Twitter or NaMo App. 
To get the one-to-one mapping of NaMo tweets with its matched Non-NaMo tweets (tweets similar to NaMo tweets):
\begin{enumerate} 
\item{We calculated the cosine distance of each NaMo tweet with Non-NaMo tweet.}
\item{We mapped a NaMo tweet to another Non-NaMo tweet with maximum similarity score.}
\end{enumerate} 

\subsubsection{2019 Lok Sabha Election Dataset}
We compared the timestamps of tweets posted using the NaMo App with its corresponding similar Non-NaMo tweet. We observed:
\begin{enumerate} 
\item{70\% times (2919/4170) NaMo tweets came first}
\item{30\% times (1251/4170) Non-NaMo tweet came first}
\end{enumerate} 

As tweets have retweets, and the same tweet can be posted more than once. We removed matching similar tweets, considering only the first instance, and performed one-to-one mapping again. We were then left with only 940 unique matched tweets.
We then observed:
\begin{enumerate} 
\item{31.4\% times (296/940) NaMo tweet came first}
\item{68.6\% times (644/940) Non NaMo tweet came first}
\end{enumerate}

\subsubsection{CAA Protests}
Similarly for the experiments done in 2019 Lok Sabha election dataset, we also performed temporal analysis for the CAA dataset.
We compared the timestamps of tweets posted using the NaMo App with its corresponding similar Non-NaMo tweet. We observed:
\begin{enumerate} 
\item{28\% times (5749/20500) NaMo tweets came first}
\item{72\% times (14751/20500) Non-NaMo tweet came first}
\end{enumerate} 

After removing matched similar tweets, considering only the first instance, and performed one-to-one mapping again. We are then left with only 2811 unique matched tweets.
We then observed:
\begin{enumerate} 
\item{2.6\% times (71/2811) NaMo tweet came first}
\item{97.4\% times (2740/2811) Non NaMo tweet came first}
\end{enumerate} 

In both cases, we observed that the tweets made using the NaMo App were already present on Twitter. No new information was coming from NaMo App. Users who use NaMo App posted content on Twitter first.

\subsection{Content Analysis}
We created word clouds to see what type of content is shared from the NaMo App to Twitter.

\begin{figure}[h]
    \begin{subfigure}[b]{0.49\textwidth}
        \begin{center}
        \includegraphics[width=50mm, height=50mm, frame]{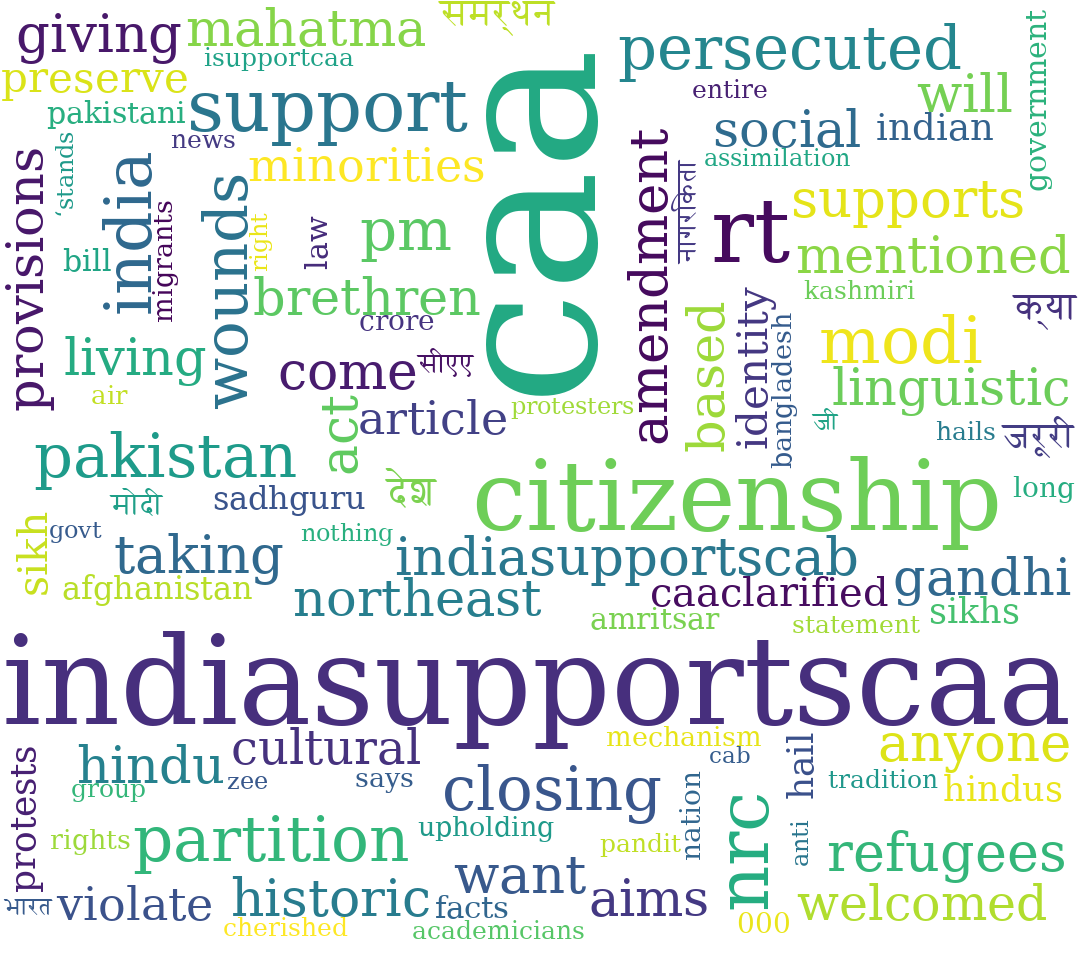}
        \subcaption{Tweets posted using NaMo App}\label{caa_namo}
        \end{center}
    \end{subfigure}
    \begin{subfigure}[b]{0.49\textwidth}
        \begin{center}
        \includegraphics[width=50mm, height=50mm, frame]{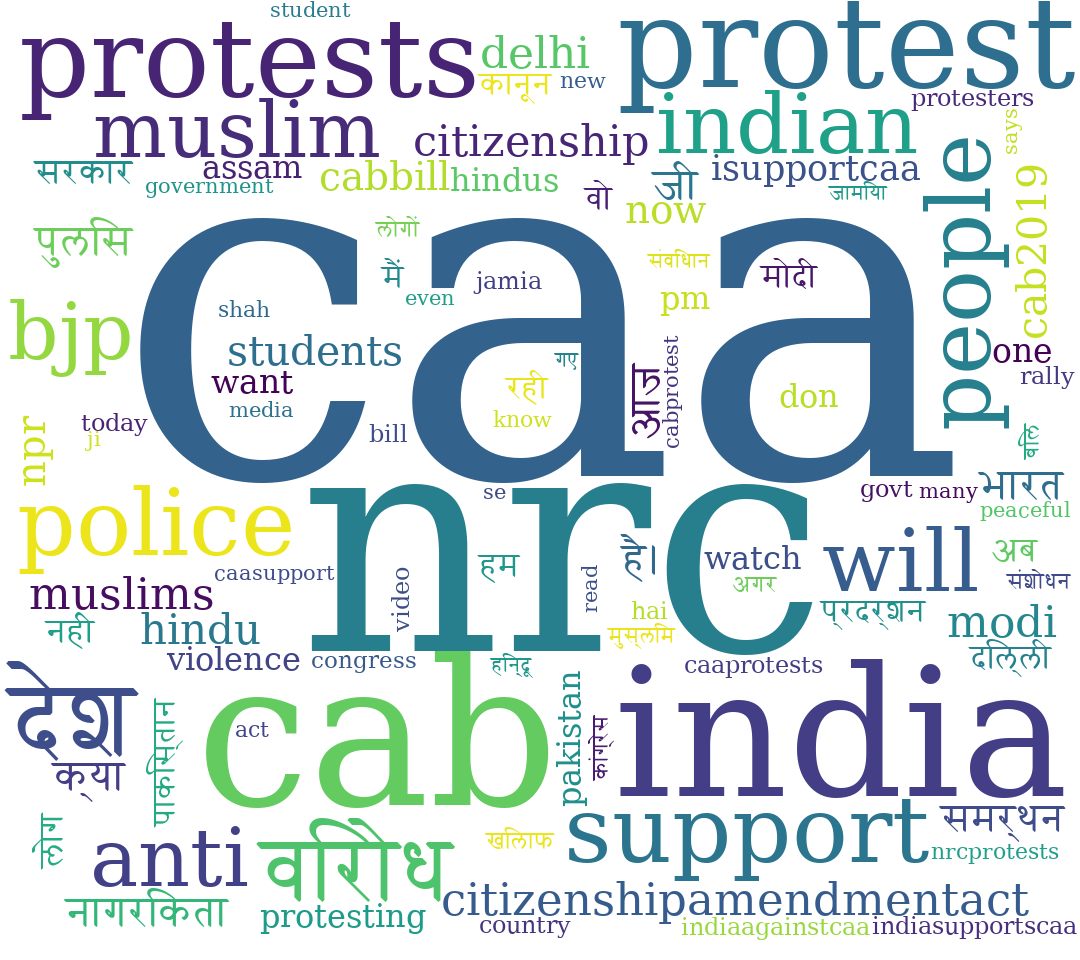}
        \subcaption{Tweets not posted using NaMo App}\label{caa_not_namo}
        \end{center}
    \end{subfigure}
    \caption{Word cloud for CAA protests.}\label{caa_n}
\end{figure} 

For the CAA dataset, we created word clouds of all the tweets shared using the NaMo App and tweets that were not posted from the NaMo App (Figure ~\ref{caa_n}). We observed that phrases such as ‘indiasupportscaa', ‘caaclarified', ‘support', and terms related to diversity or places such as ‘pakistan', ‘northeast', ‘amritsar', ‘afghanistan, ‘sikhs', etc., are predominent in the word cloud of the tweets posted using NaMo App (Figure ~\ref{caa_namo}). Whereas in the word cloud of the tweets not posted using the NaMo App (Figure ~\ref{caa_not_namo}), phrases such as ‘protests', ‘indiaagainstcaa', and ‘anti' are present. \\

\begin{figure}[h]
    \begin{subfigure}[b]{0.49\textwidth}
        \begin{center}
        \includegraphics[width=50mm, height=50mm, frame]{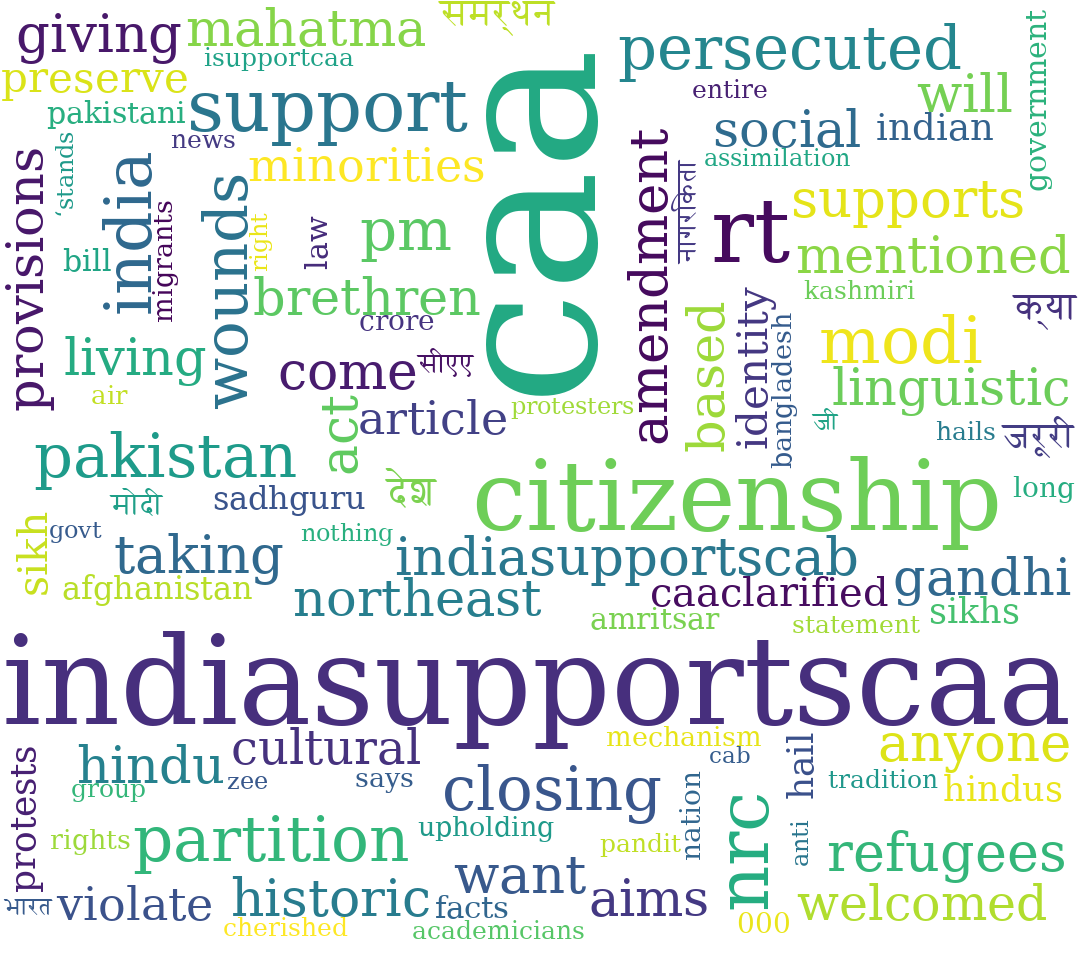}
        \subcaption{Tweets posted using NaMo App}\label{nn2}
        \end{center}
    \end{subfigure}
    \begin{subfigure}[b]{0.49\textwidth}
        \begin{center}
        \includegraphics[width=50mm, height=50mm, frame]{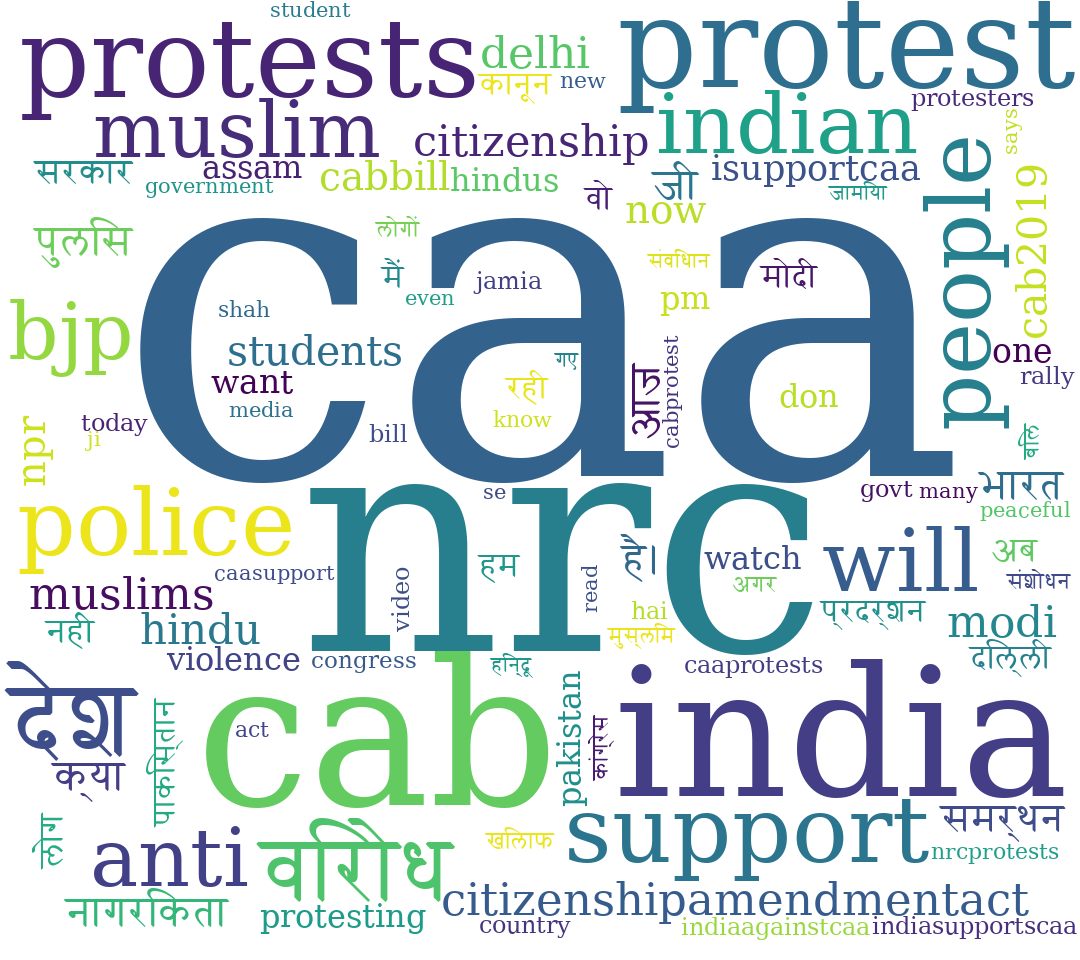}
        \subcaption{Tweets not posted using NaMo App}\label{nn}
        \end{center}
    \end{subfigure}
    \caption{Word cloud for seed users for 2019 Lok Sabha Election.}\label{election_seed}
\end{figure} 

For the 2019 Lok Sabha Election dataset, we created word clouds for the seed users and auxiliary users. Word cloud of tweets posted using NaMo App by seed users (Figure ~\ref{nn2}) contains pro-BJP phrases such as ‘deshkagauravmodi', ‘indiavotesfornamo', ‘aayegatomodihi', ‘indiavotesfornamo' and other terms related to positivity or BJP schemes such as ‘benefits',  ‘employment', ‘dhan', ‘aadhar', ‘transparent', ‘opportunities', etc., are also present. Whereas in word cloud (Figure ~\ref{nn}) of tweets not posted using NaMo App, we observe terms related to other parties also such as ‘gandhi', ‘congress', ‘{\dn mmtA}', ‘{\dn Ed\3A5wF}', ‘{\dn kAfF }', ‘varanasi', ‘{\dn p=\8{p}}', etc.

\begin{figure}[H]
    \begin{subfigure}[b]{0.49\textwidth}
        \begin{center}
        \includegraphics[width=50mm, height=50mm, frame]{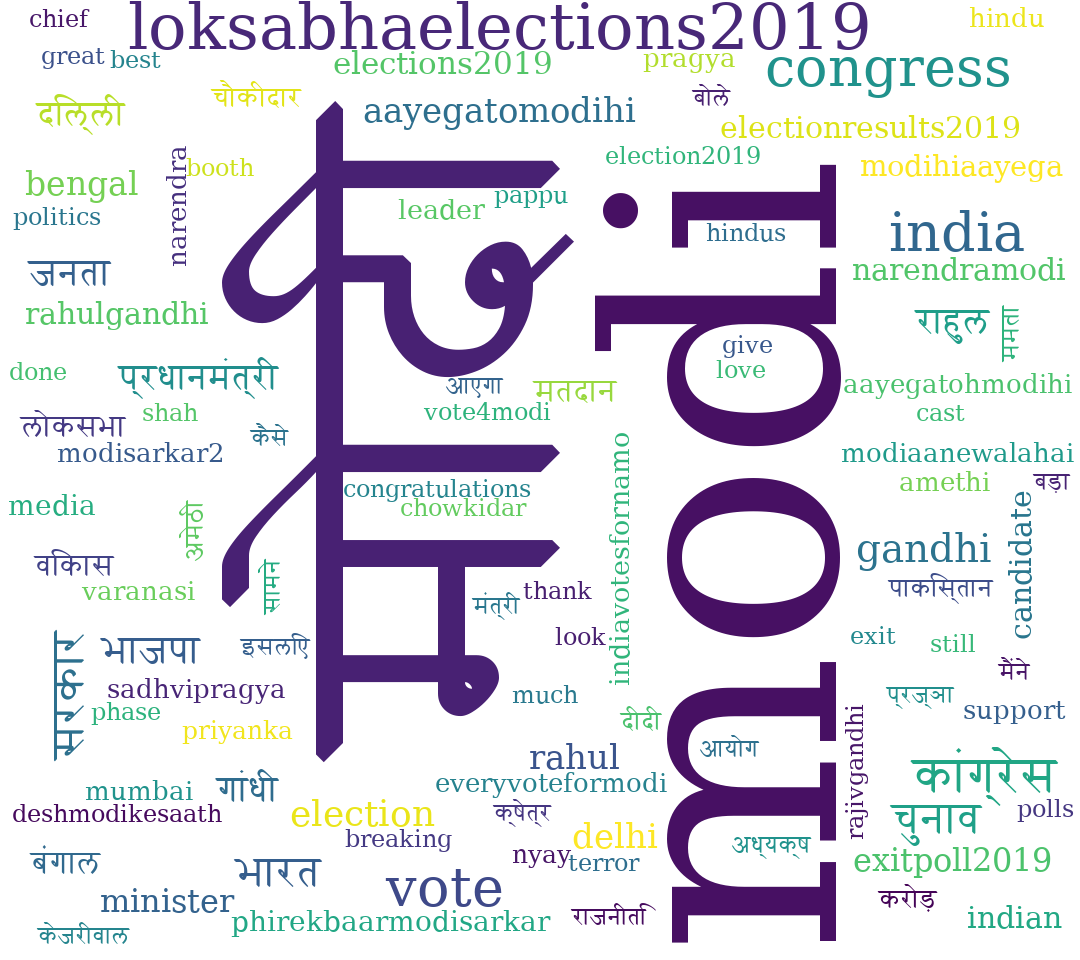}
        \subcaption{Tweets similar to NaMo tweets}\label{fig:test3}
        \end{center}
    \end{subfigure}
    \begin{subfigure}[b]{0.49\textwidth}
        \begin{center}
        \includegraphics[width=50mm, height=50mm, frame]{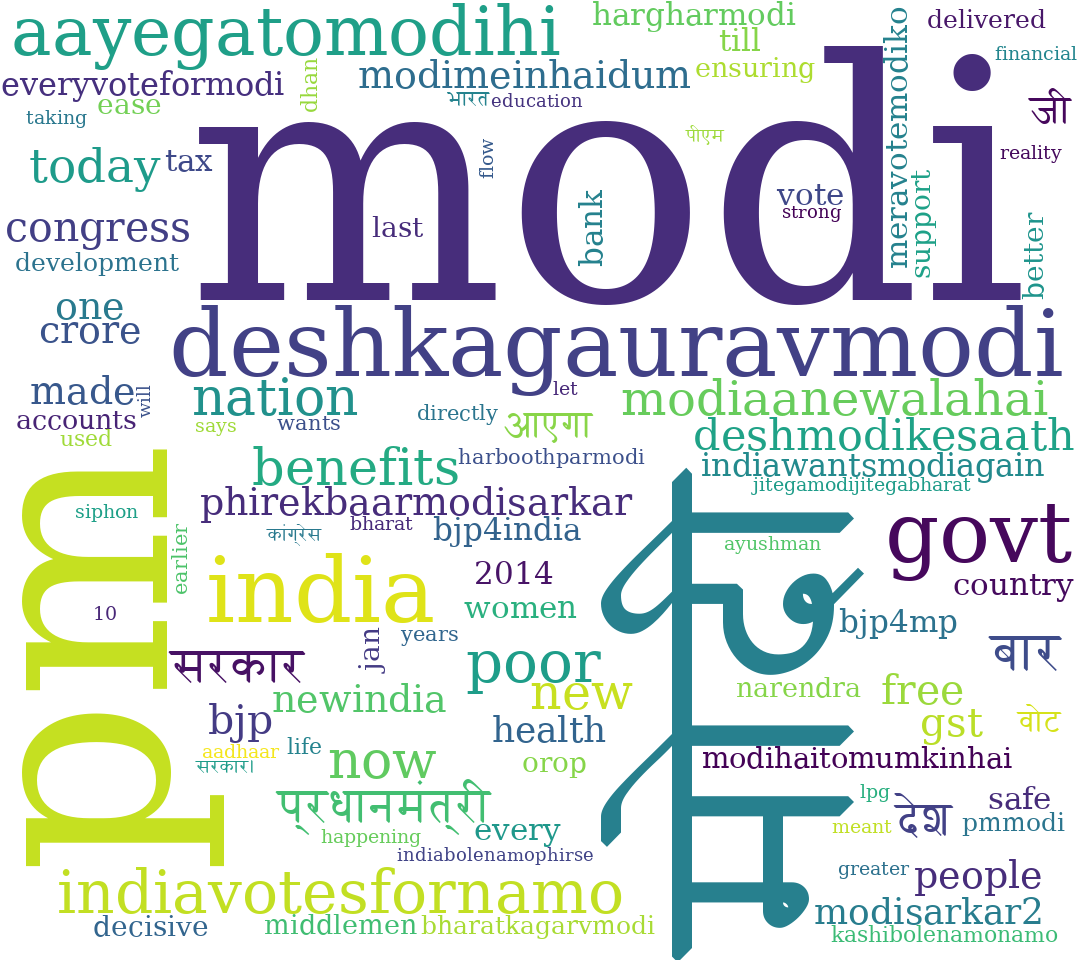}
        \subcaption{Tweets not similar to NaMo tweets}\label{fig:test4}
        \end{center}
    \end{subfigure}
    \caption{Word cloud for auxiliary users for 2019 Lok Sabha Election.}\label{election_aux}
\end{figure}

In the word cloud of auxiliary users (Figure ~\ref{election_aux}), all the highly pro-BJP hashtags were present such as ‘deshkagauravmodi', ‘modiaanewalahai', ‘indiawantsmodiagain', ‘aayegatomodihi', ‘modimeihaidum', ‘phirekbaarmodisarkar', etc. While other terms have less frequencies. Through all the word clouds, we can conclude that users who uses NaMo App to post content or users who post content similar to NaMo tweets are highly pro-BJP. 

%% file: Chapters/Chapter7.tex

\section{Hawkes Process} 

\label{Hawkes Process} 

To answer our RQ2 - How much influence does NaMo content have on Twitter? - We utilized the Hawkes process \cite{26}. Some of the terminologies are: 
\begin{enumerate}
\item{\textbf{Point Processes:} Point processes are collections of random points falling in some space, such as time and location. It describes the timing and properties of events.}
\item{\textbf{Poisson Point Process:} If an event does not impact the rate of future occurrence of other events, then it is called a poisson point process. They are memoryless means the inter-arrival between events depends only on relevant information about the current time, but not on information from further in the past. Events arrive randomly with the constant intensity $\lambda$.}
\item{\textbf{Self-Excited Process:} If an event increases the rate of future occurrence of other events, then it can be categorized as self-excited process. They are not memoryless in nature.}
\end{enumerate}
Hawkes Processes are self-exciting temporal point processes that describe how events occur on a set of processes.  A Hawkes model consists of K number of point processes, each with a ‘background rate', which determines the rate at which events occur in that process without any external influence. However, an event on one process can cause a response on other processes, which increases the probability of an event occurring on other processes.\\
Hawkes process can be used for modeling the information flow in web communities as well as for measuring social influence \cite{20,21,22,23,24,25}. As web communities do not exist in a vacuum, there is a probability that if an image is posted on one web community, it gets posted on another web community \cite{22}. In our setting, events are the posting of an image, and processes are web communities.  Each web community has its own background rates for posting the images as well as some influence due to other social media sites. For our purpose, fitting a Hawkes model gives us values for the background rates for each platform as well as the likelihood or probability of an event on one platform triggering events on other platforms.\\
For a Hawkes model, the rate of each k-th process,
$\lambda$ $_{t,k}$ is given by:

\[
    \lambda_{t,k} = \lambda_{0,k} + 
    {\sum_{k^{'}=1}^{K}}
    {\sum_{t^{'}=1}^{t-1}} s_{t^{'},k^{'}} \times h_{k^{'}\rightarrow k}[t−t^{'}]
\]
where $s$ $\in$ N$^{T\times K}$ is the matrix of event counts and $h_{k^{'}\rightarrow k}[t−t^{'}]$ is an impulse response
function that gives amount of influence that process k$^{'}$ have on the rate of process k. The impulse response function $h_{k^{'}\rightarrow k}[t−t^{'}]$ can be decomposed into a scalar weight $W_{k^{'}\rightarrow k}$ and a probability
mass function $G_{k^{'}\rightarrow k}[d]$  \cite{56}. The weight describes the intensity of the interaction from process k to process k$^{'}$ and the probability mass function specifies how the interaction changes over time:
\[
    h_{k^{'}\rightarrow k}[d] = W_{k\rightarrow k^{'}} \times G_{k\rightarrow k^{'}}[d]
\]

According to Zannettou et al. \cite{22},the weight value $W_{k\rightarrow k^{'}}$ can be interpreted as the expected number of events caused on process $k^{'}$ due to an event on process k which gives the influence that one web community have on another web community. The probability mass function $G_{k\rightarrow k^{'}}$ specifies the probability that a child event will occur at each specific time lag d\( \Delta t \), up to a maximum lag  \( \Delta t \)$_{max}$.

\subsection{Methodology}
For our research, we fitted the Hawkes model as described in \cite{22} with K as 2, i.e., the number of platforms - Twitter and NaMo App - for each image posted on both the platforms. 
\begin{enumerate}
\item{ We calculated perceptual hash (pHash) \cite{33} for each image posted during CAA protests on both the platforms using the ImageHash library\footnote{\url{https://pypi.org/project/ImageHash/}}.}
\item{ We filtered out the images on the NaMo App outside the timeline of the Twitter dataset of CAA protests. Number of images present in the Twitter dataset of CAA dataset - 1,68,974 and images present on NaMo App during CAA protests - 1,04,737.}
\item{ We filtered out the hashes that were not present in both the platforms, i.e., the images not posted on both the platforms. After filtering, we found 2,168 images that are common on both platforms.}
\item{ For each image hash, we created a matrix
$s$ $\in$ N$^{T\times 2}$ containing the number of events of a particular image posting per minute for each of the platforms/subreddits. Here, T is the number of
minutes from the first recorded post of the image hash on any platform to
the last recorded post of an image hash on another platform.}
\item{ Hawkes model also takes one more parameter \( \Delta t \) that defines a given event on one platform can cause an event on other platforms within the given time window. We performed the Hawkes process for three different values of \( \Delta t \) - 12 hours, 24 hours, and 48 hours -  each converted into minutes.}
\item{ Fitting Hawkes model gives us the values of weight matrix W$^{2\times 2}$ - which amounts to the influence between the two platforms for each image. The weight value act as a rate parameter in the Poisson process. It can be interpreted as the expected number of events on a platform due to another platform. }
\item{ After obtaining the weight matrix W$^{2\times 2}$ of each image hash and for each \( \Delta t \), we added up all the matrices to calculate the mean weight matrix (Table ~\ref{tab:freq1}, ~\ref{tab:freq2}, ~\ref{tab:freq3}). }
\end{enumerate}

\begin{table}[!htb]
  
  \begin{tabular}{ccl}
    \toprule
    \textbf{} & \textbf{NaMo App} & \textbf{Twitter} \\
    \midrule
    \bf NaMo App &  0.25619469 & 0.13528167 \\
     \bf Twitter &  0.1804032 & 0.2548587\\
  \bottomrule
\end{tabular}
\caption{Weight matrix obtained through Hawkes process. (For 12 hour time period)}
\label{tab:freq1}
\end{table}

\begin{table}[!htb]
 
  \begin{tabular}{ccl}
    \toprule
    \textbf{} & \textbf{NaMo App} & \textbf{Twitter} \\
    \midrule
    \bf NaMo App &  0.25667107 & 0.13347803 \\
    \bf Twitter &  0.18027061 & 0.25673516  \\
  \bottomrule
\end{tabular}
\caption{Weight matrix obtained through Hawkes process. (For 24 hour time period)}
 \label{tab:freq2}
\end{table}

\begin{table}[!htb]
  
  \begin{tabular}{ccl}
    \toprule
    \textbf{} & \textbf{NaMo App} & \textbf{Twitter} \\
    \midrule
    \bf NaMo App & 0.25685813 & 0.1402505 \\
     \bf Twitter &  0.54783613 & 0.25523964 \\
  \bottomrule
\end{tabular}
\caption{Weight matrix obtained through Hawkes process. (For 48 hour time period)}
\label{tab:freq3}
\end{table}

\subsection{Observations}
We used Hawkes process because it allows us to quantify the effect of Web communities while also taking into consideration the influence of other information sources. Hawkes process can be applied to any number of web communities. We applied Hawkes on NaMo and Twitter to check their influence on each other. Through weighted matrices ( ~\ref{tab:freq1}, ~\ref{tab:freq2}, ~\ref{tab:freq3} ), we observed similar values for all the time periods. Retweeting activity on twitter can be attributed to the same image being shared on the same platform, whereas for the NaMo App, the value can be used to indicate that the same image is posted by some other user. However, the influence that Twitter has on the NaMo App is greater than NaMo App to Twitter, indicating that the content is first shared on Twitter and then circulated or posted to NaMo App through other means. The results are similar to the results we obtained through temporal analysis, that the content similar to the NaMo posts posted on twitter, is posted first.

%% file: Chapters/Chapter8.tex

\section{Characterising the users that post
NaMo App content on Twitter} 

\label{Characterising the users that post
NaMo App content on Twitter} 

To answer the RQ3 - Can we characterize the users that post NaMo content on Twitter? - We followed the following methodology:

\subsection {Are users part of an Echo Chamber on Twitter?}

\subsubsection{Methodology}
We assigned a score of 0 or 1 to each hashtag for all the users that posts NaMo App content on twitter as described in Chapter ~\ref{Echo Chamber during Indian Elections} . An anti-BJP hashtag or a pro-Congress hashtag is assigned a score of 0, whereas a pro-BJP hashtag is assigned a score of 1. We calculated the values of production polarity, production variance, consumption polarity, and consumption variance for every seed and auxiliary user in our database as described in Chapter ~\ref{Echo Chamber during Indian Elections}.

\subsubsection{Observations}
\begin{figure}[H]
    \begin{subfigure}[b]{0.49\textwidth}
        \begin{center}
        \includegraphics[width=60mm, height=50mm]{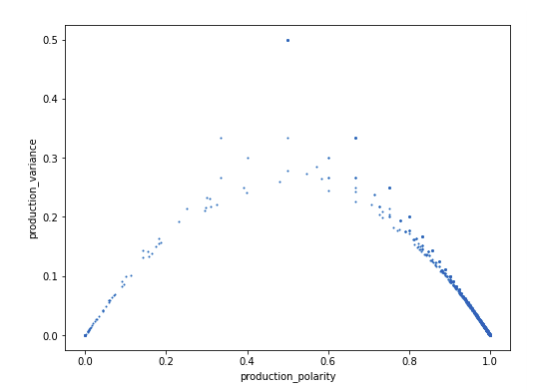}
        \subcaption{Production Variance vs Production Polarity}\label{8}
        \end{center}
    \end{subfigure}
    \begin{subfigure}[b]{0.49\textwidth}
        \begin{center}
        \includegraphics[width=60mm, height=50mm]{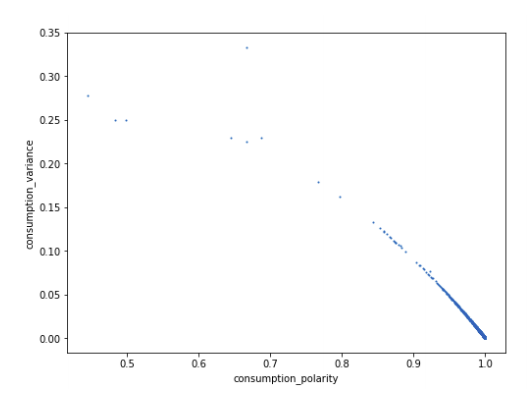}
        \subcaption{Consumption Variance vs Consumption Polarity}\label{9}
        \end{center}
    \end{subfigure}
    \caption{Graph for Variance vs Polarity.}\label{bjp}
\end{figure}

\begin{figure}[h]  
\centering
\includegraphics[width=0.6\linewidth]{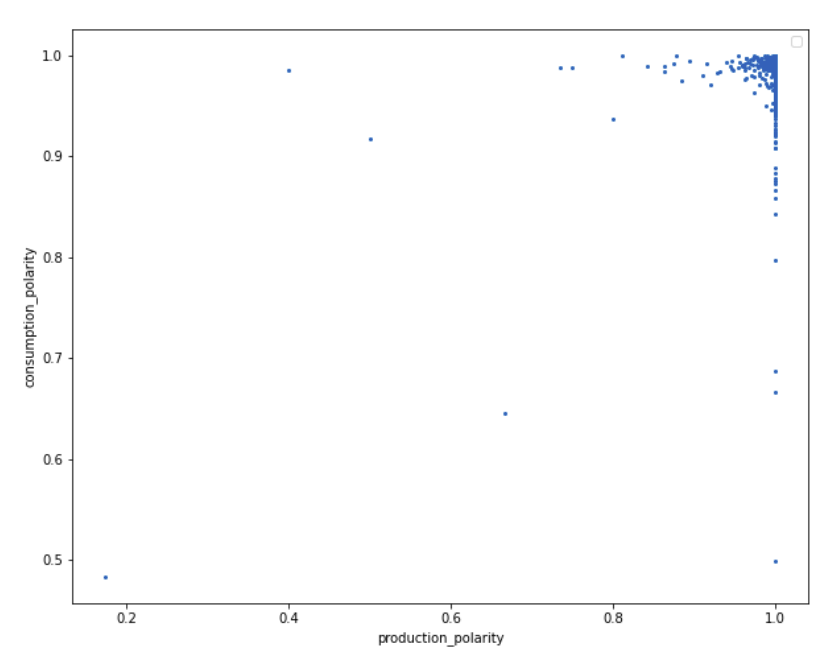}
\caption{Consumption Polarity vs Production Polarity.}\label{copo}
\end{figure}

This section aims to see the affiliation of users that post tweets using the NaMo App. We performed experiments to show the presence of echo chambers as outlined by Garimella et al. \cite{5}. We examine three plots for each user - production polarity vs. production variance (Figure ~\ref{8}), consumption polarity vs. consumption variance (Figure ~\ref{9}), and production polarity vs. consumption polarity (Figure ~\ref{copo}).  According to our findings in (Figure ~\ref{8}) , production variance for users at the extremes of production polarity tends to be close to zero. In contrast, users with production polarity close to 0.5 have the highest production variance - leading to a downward U shape - showing that the users who post content of a particular ideology have low variation in the content they produce. The users with polarity values close to 0.5 are bipartisan, i.e., users who consume news with a wider spread of political leaning, rather than just a particular ideology, making their news diet different from partisan users. We observed that most users have production polarity closer to 1, indicating that they are more likely to create tweets that favor the BJP.

In Figure ~\ref{9}, we observed same results as that of production polarity and production variance. The users who consume content of average polarity close to 1 have minor variation in the polarity of their consumed tweets. Also, most users have consumption polarity close to 1, indicating that they tend to consume tweets that support BJP. 
We also plotted the production polarity and consumption polarity scores for each user in Figure ~\ref{copo}. At extreme ends of the plot, the users are partisan in nature, i.e., they are producing and consuming content from a narrower spectrum of sources. We see that the majority of the users are clustered in the plot's upper-right corner of the graph. As pro-Congress and anti-BJP hashtags have a score of 0 and pro-BJP hashtags have a score of 1, the plot reveals that most people who post NaMo content on Twitter generate and consume tweets that reflect their pro-BJP ideology.

\subsection { Do users of certain states form a larger part of affected user
group?}
\subsubsection{Methodology}
In order to obtain the location for Twitter users, we used place object\footnote{\url{ https://developer.twitter.com/en/docs/twitter-api/v1/data-dictionary/object-model/geo}} and user object\footnote{\url{ https://developer.twitter.com/en/docs/twitter-api/v1/data-dictionary/overview/user-object}}. We followed the steps described below:\\

\begin{figure}[H]
    \begin{subfigure}[b]{0.49\textwidth}
        \begin{center}
        \includegraphics[width=60mm, height=50mm]{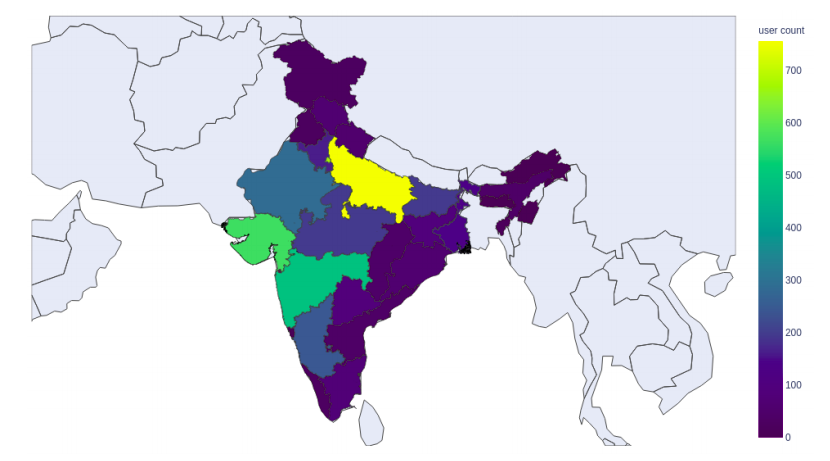}
        \subcaption{For 2019 Lok Sabha Elections}\label{10}
        \end{center}
    \end{subfigure}
    \begin{subfigure}[b]{0.49\textwidth}
        \begin{center}
        \includegraphics[width=60mm, height=50mm]{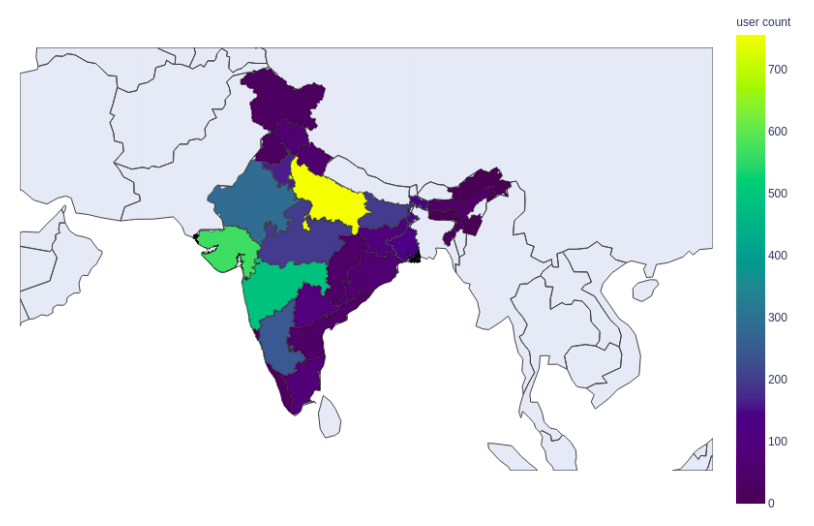}
        \subcaption{For CAA Protests}\label{11}
        \end{center}
    \end{subfigure}
    \caption{State-wise Heat map of the count of Affected Users according to their location.}\label{4bjp}
\end{figure}

\begin{lstlisting}[
    basicstyle=\tiny
]
If user had geo-tagging turned on :
    location  = place_object['full_name']
else :
    location = user_object['location']

if location contains names of multiple cities OR foreign cities OR foreign countries:
    Remove user from the database
else:
    if location is an Indian city:
        location = mapping_of_city_to_state[city_name]
\end{lstlisting}
After mapping, we plotted heat maps of the number of Twitter users belonging to a given state (Figure ~\ref{10}, Figure ~\ref{11}).

\subsubsection{Observations}
We observed that Uttar Pradesh had the most affected users in both the 2019 Lok Sabha Election and the CAA Protest, followed by the NCT of Delhi and Gujarat. When comparing affected users to general users (Table ~\ref{t1}), we observed that the states most closely associated with pro-BJP views - Uttar Pradesh and Gujarat - show a significant rise in Twitter users. In Gujarat, for example, we noticed that the proportion of affected users got doubled.

\begin{table}[!htb]

  \begin{tabular}{ccl}
    \toprule
    \textbf{State } & \textbf{Fraction of affected
users} & \textbf{Fraction of general users}\\
    \midrule
    
 Uttar Pradesh & 0.171811 & 0.140201 \\
  \hline
NCT of Delhi & 0.156173 & 0.184110 \\
 \hline
Gujarat & 0.128189 & 0.057137 \\
 \hline
Maharashtra & 0.109671 & 0.149511 \\
 \hline
Rajasthan & 0.062963 & 0.057609 \\
  \bottomrule
\end{tabular}
\caption{Fraction of users (for 2019 Lok Sabha Elections dataset)}
\label{t1}
\end{table}

\subsection {Is there a difference between Seed users and Auxiliary users?}

\subsubsection{Methodology}
\begin{enumerate}
  
\item {We converted the Twitter user descriptions into lowercase. We removed all punctuation, Twitter handles, hashtags, emojis, links, and stopwords.}
\item {We created a word cloud of the tweets of the seed users and auxiliary users.}
\item {We also calculated odds ratios \cite{34} for the description of affected users. The odds ratio is a statistical metric that quantifies `the odds that an outcome will occur given a particular exposure, compared to the odds of the outcome occurring in the absence of that exposure'. We measured the odds ratios of all the bigrams and trigrams for the seed and auxiliary user descriptions. To get the n-grams, we used CountVectorizer\footnote{\url{
https://scikit-learn.org/stable/modules/generated/sklearn.feature\_extraction.text.CountVectorizer.html}} with Indic tokenize\footnote{\url{
https://indic-nlp-library.readthedocs.io/en/latest/\_modules/indicnlp/tokenize/indic\_
tokenize.html}} as the tokenizer. A higher odds ratio for a given bigram or trigram for a given class indicates that the bigram or trigram has a close relationship with that class.}
\end{enumerate}

\begin{figure}[h]
    \begin{subfigure}[b]{0.49\textwidth}
        \begin{center}
        \includegraphics[width=60mm, height=50mm, frame]{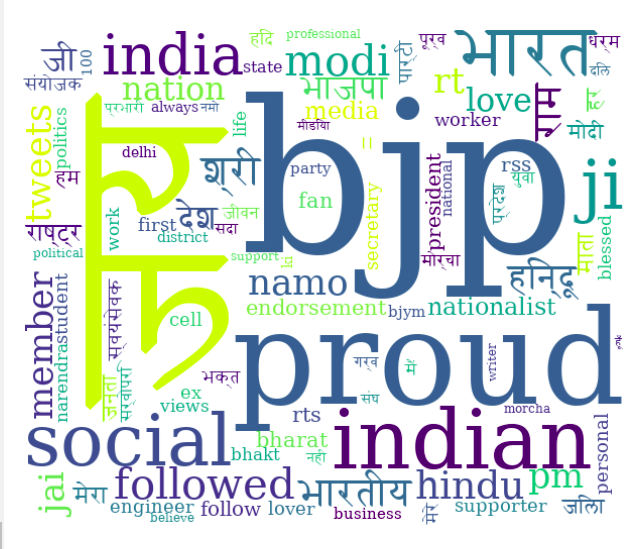}
        \subcaption{For seed users}\label{101}
        \end{center}
    \end{subfigure}
    \begin{subfigure}[b]{0.49\textwidth}
        \begin{center}
        \includegraphics[width=60mm, height=50mm, frame]{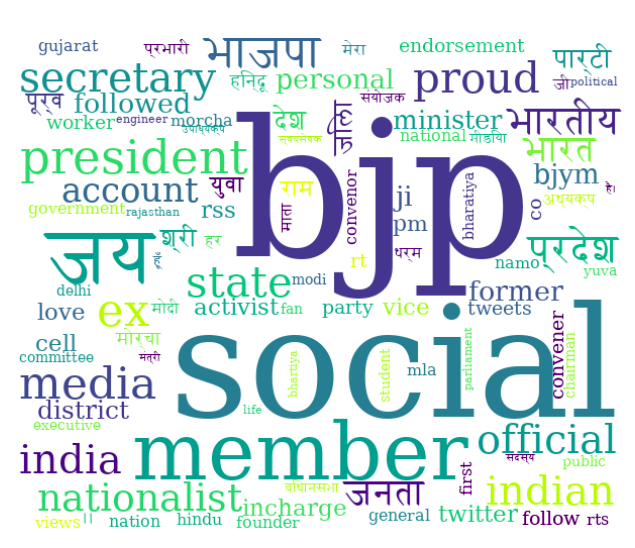}
        \subcaption{For auxiliary users}\label{102}
        \end{center}
    \end{subfigure}
    \caption{Word cloud of user descriptions (For 2019 Lok Sabha Elections)}\label{wc}
\end{figure}

\begin{figure}[h]
    \begin{subfigure}[b]{0.49\textwidth}
        \begin{center}
        \includegraphics[width=60mm, height=50mm, frame]{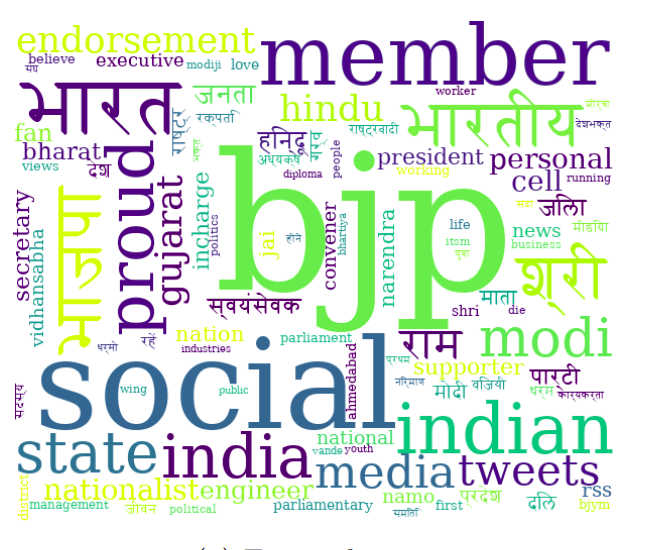}
        \subcaption{For seed users}\label{fig:test3}
        \end{center}
    \end{subfigure}
    \begin{subfigure}[b]{0.49\textwidth}
        \begin{center}
        \includegraphics[width=60mm, height=50mm, frame]{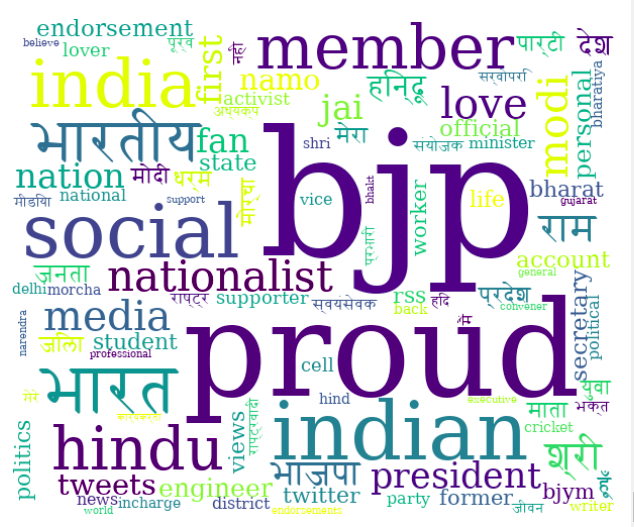}
        \subcaption{For auxiliary users}\label{fig:test4}
        \end{center}
    \end{subfigure}
    \caption{Word cloud of user descriptions (For CAA Protests)}\label{wc2}
\end{figure} 

\subsubsection{Observations}
\textbf{For 2019 Lok Sabha Elections:}\\
In the word cloud of user descriptions of 2019 Lok Sabha election (Figure ~\ref{wc}), ‘bjp' is one of the most frequently occurring word for both seed and auxiliary users. \\
\textbf{For seed users} - we observed that the odd ratios are high for ngrams including phrases such as ‘social media soldiers', ‘{\dn aAIVF s\2yojk}', ‘{\dn -vy\2 s\?vk BAjpA kAy\0ktA\0}', ‘{\dn EjlA \3FEw\7{m}K aAIVF}', `{\dn rA\3A3wvAdF{\qvb} nmo B\3C4w}' - indicating that seed users possibly belongs to BJP IT cell or are grass-root supporters or workers of BJP party. \\
\textbf{For auxiliary users} - we observed terms related to positions of power such as ‘secretary', ‘president', ‘minister', ‘incharge', ‘convener', etc., present in the word cloud for auxiliary users but not in seed users. We also observed that the values of the odd-ratios are high for ngrams such as ‘minister state',  ‘president byjm', ‘union member', ‘member legislative assembly', ‘committee member' etc. One might expect to see such phrases in the profile bio of a prominent user. This suggests that auxiliary users are likely to be associated with influential users within the BJP party and its affiliated groups. In contrast, seed users are likely to be BJP followers or lower-level members/workers of BJP party. As seed users utilises the NaMo App to post content to Twitter, whereas auxiliary users post the same content directly to Twitter, it is logical that they would already have access to the content owing to association with influential users and would not need to utilise the NaMo App to cross-post. On the other hand, lower-level workers or members would only have access to such information through the NaMo App, making them part of the seed user group.

\begin{figure}[H]
    \begin{subfigure}[b]{0.49\textwidth}
        \begin{center}
        \includegraphics[width=40mm, height=80mm]{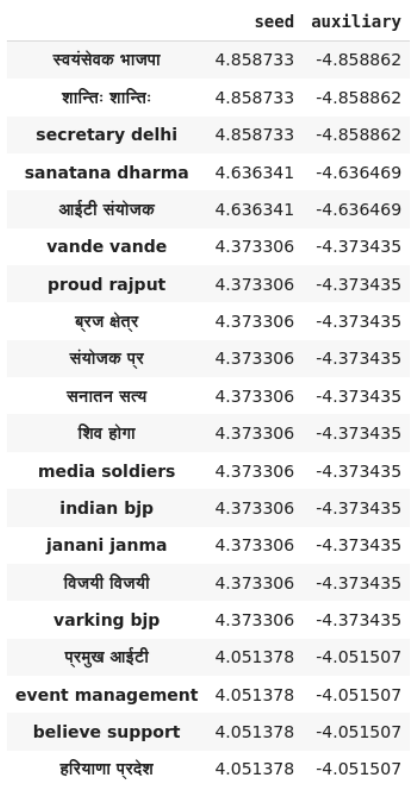}
        \subcaption{For seed users}\label{fig:a}
        \end{center}
    \end{subfigure}
    \begin{subfigure}[b]{0.49\textwidth}
        \begin{center}
        \includegraphics[width=40mm, height=80mm]{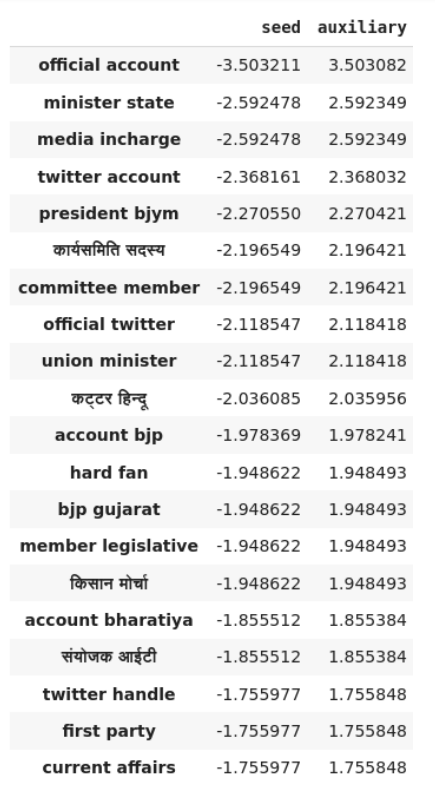}
        \subcaption{For auxiliary users}\label{fig:b}
        \end{center}
    \end{subfigure}
    \caption{Odd-ratios of bigrams of user descriptions (For 2019 Lok Sabha Elections).}\label{figure}
\end{figure}

\begin{figure}[H]
    \begin{subfigure}[b]{0.49\textwidth}
        \begin{center}
        \includegraphics[width=40mm, height=80mm]{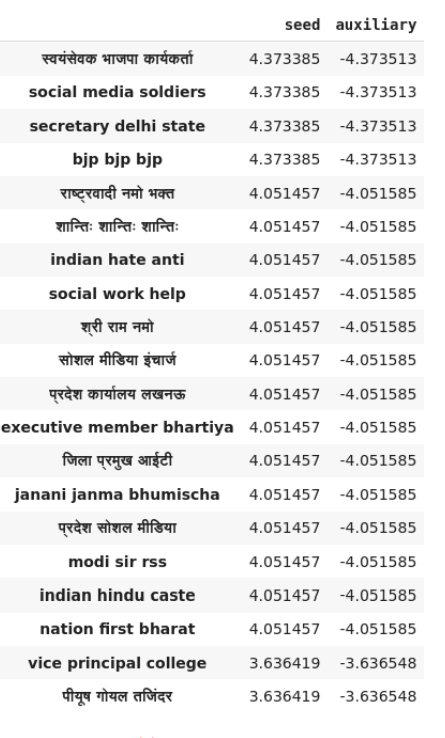}
        \subcaption{For seed users}\label{fig:testa}
        \end{center}
    \end{subfigure}
    \begin{subfigure}[b]{0.49\textwidth}
        \begin{center}
        \includegraphics[width=40mm, height=80mm]{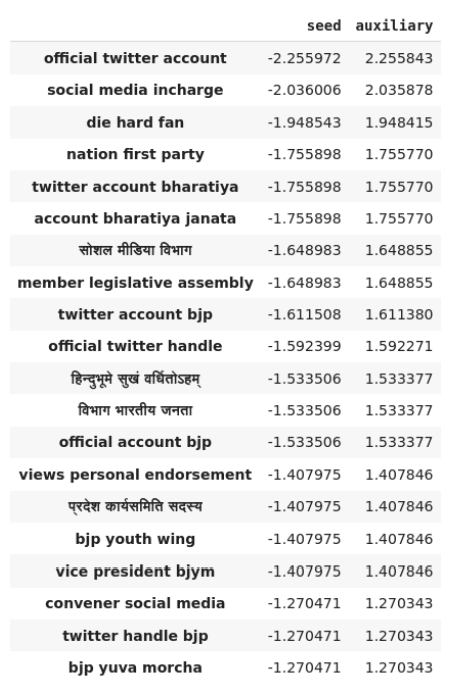}
        \subcaption{For auxiliary users}\label{fig:testb}
        \end{center}
    \end{subfigure}
    \caption{Odd-ratios of trigrams of user descriptions (For 2019 Lok Sabha Elections).}\label{figure}
\end{figure} 

\textbf{For CAA Protests:}\\
We observed similar results for CAA protests dataset. Words like ‘secretary', ‘president', ‘minister', ‘member', ‘convener', ‘incharge', etc., appeared in both of the word clouds of twitter descriptions of seed and auxiliary users.\\
\textbf{For seed users} - we observed odd-ratios were high for phrases such as ‘{\dn -vys\2 s\?vk rA\6{\3A3w}B\3C4w} bjym', ‘{\dn -vy\2 s\?vk BAjpA kAy\0ktA\0}', ‘{\dn B\3C4w d\?fB\3C4w}', `{\dn \8{h} spoV\0r}' - suggesting to the possibility of seed users to be grass-root supporters or workers of organizations such as BJP, RSS. \\
\textbf{For auxiliary users} - we observed that auxiliary users have high odd-ratios of bigrams and trigrams for phrases like ‘official twitter account', ‘vice president', ‘official twitter handle', ‘incharge social media', etc. \\
Thus, we can infer from both datasets that auxiliary users possibly hold close relations with users holding positions in the BJP party. In contrast, seed users are likely to be supporters or lower-level members of BJP and associated organizations.

\begin{figure}[H]
    \begin{subfigure}[b]{0.49\textwidth}
        \begin{center}
        \includegraphics[width=40mm, height=80mm]{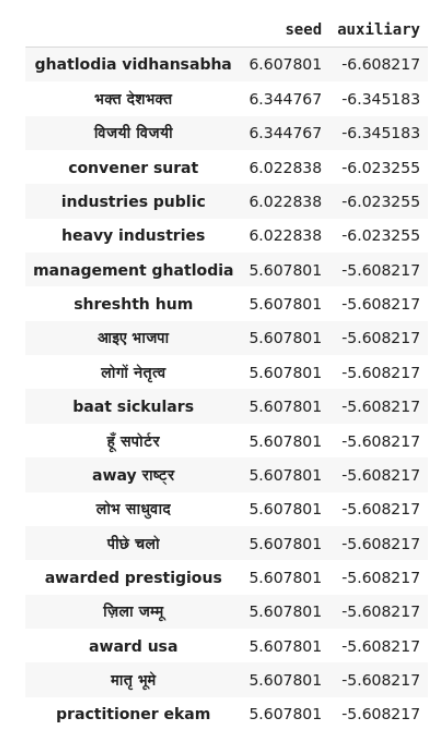}
        \subcaption{For seed users}\label{fig:a}
        \end{center}
    \end{subfigure}
    \begin{subfigure}[b]{0.49\textwidth}
        \begin{center}
        \includegraphics[width=40mm, height=80mm]{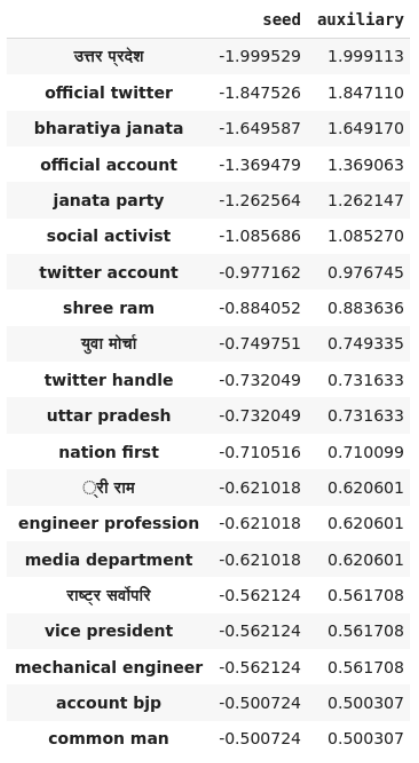}
        \subcaption{For auxiliary users}\label{fig:b}
        \end{center}
    \end{subfigure}
    \caption{Odd-ratios of bigrams of user descriptions (For CAA Protests).}\label{}
\end{figure}

\begin{figure}[H]
    \begin{subfigure}[b]{0.49\textwidth}
        \begin{center}
        \includegraphics[width=40mm, height=80mm]{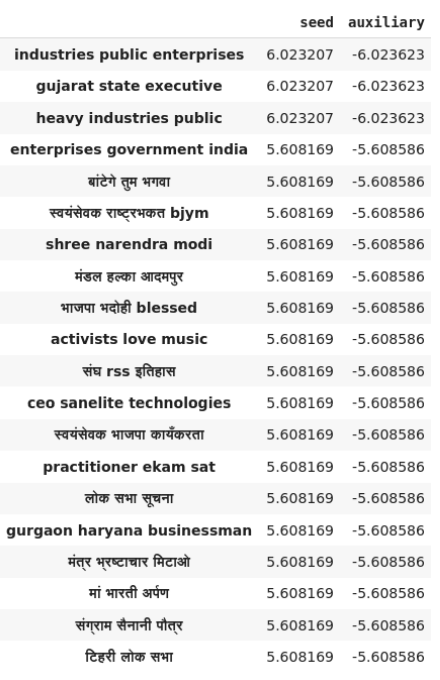}
        \subcaption{For seed users}\label{fig:test3}
        \end{center}
    \end{subfigure}
    \begin{subfigure}[b]{0.49\textwidth}
        \begin{center}
        \includegraphics[width=40mm, height=80mm]{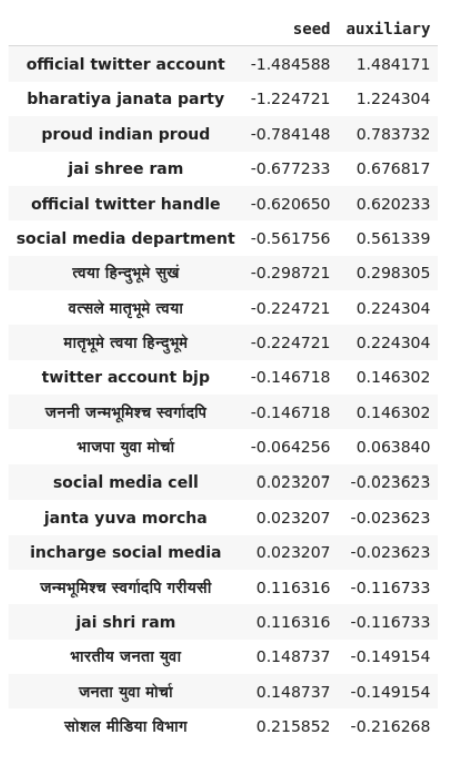}
        \subcaption{For auxiliary users}\label{fig:test4}
        \end{center}
    \end{subfigure}
    \caption{Odd-ratios of trigrams of user descriptions (For CAA Protests).}\label{}
\end{figure} 

%% file: Chapters/Chapter9.tex

\section{Reachability of the NaMo App
content on Twitter} 

\label{Reachability of the NaMo App
content on Twitter} 

To answer the RQ4 - What is NaMo content's reachability on Twitter compared to non-NaMo content? -  We created correlation graphs for likes and retweets of tweets posted from the NaMo App and the tweets that were similar to tweets posted using NaMo content.

\subsection{Methodology}
After applying K-means clustering on tweets posted from the NaMo App on Twitter as defined in chapter ~\ref{Contribution of the NaMo App to
Twitter content}, we identified Non-NaMo tweets, i.e., tweets without tags ‘via MyNt' or ‘via NaMo App' at the end.
\begin{enumerate}   
\item{For each Non-NaMo tweet, we calculated the cosine similarity with every NaMo tweet.}
\item{We map each Non-NaMo tweet with a NaMo tweet with the highest similarity.}
\item{We created the scatter plots between the number of likes and retweets for each tweet posted using NaMo App and their corresponding Non-NaMo tweet.}
\item{We then created the scatter plots after normalizing the likes and retweets with the number of user's followers so that the graph is not biased as more followed user's tweets get more likes and retweets when compared with a less followed user.}
\item {We also created CDF graphs of likes and retweets of NaMo tweets and their corresponding mapped Non-NaMo tweet.}
\end{enumerate}

\begin{figure}[h]
    \begin{subfigure}[b]{0.49\textwidth}
        \begin{center}
        \includegraphics[width=60mm, height=50mm]{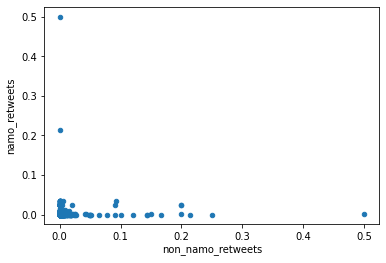}
        \subcaption{Plot for retweets}\label{fig:testa}
        \end{center}
    \end{subfigure}
    \begin{subfigure}[b]{0.49\textwidth}
        \begin{center}
        \includegraphics[width=60mm, height=50mm]{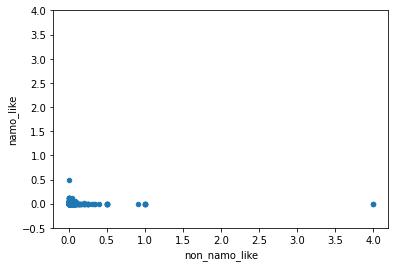}
        \subcaption{Plot for likes}\label{fig:testb}
        \end{center}
    \end{subfigure}
    \caption{Correlation graphs normalised by followers count (For 2019 Lok Sabha Elections).}\label{}
\end{figure}

\begin{figure}[h]
    \begin{subfigure}[b]{0.49\textwidth}
        \begin{center}
        \includegraphics[width=60mm, height=50mm]{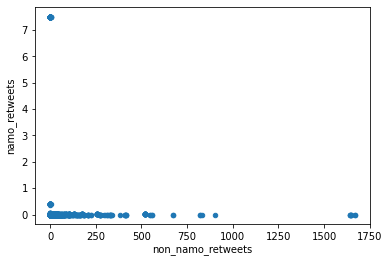}
        \subcaption{Plot for retweets}\label{fig:testa}
        \end{center}
    \end{subfigure}
    \begin{subfigure}[b]{0.49\textwidth}
        \begin{center}
        \includegraphics[width=60mm, height=50mm]{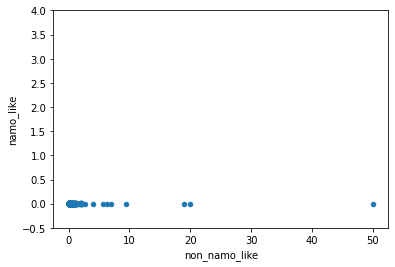}
        \subcaption{Plot for likes}\label{fig:testb}
        \end{center}
    \end{subfigure}
    \caption{Correlation graphs normalised by followers count (For CAA Protests).}\label{}
\end{figure}

\begin{figure}[h]
    \begin{subfigure}[b]{0.49\textwidth}
        \begin{center}
        \includegraphics[width=60mm, height=50mm]{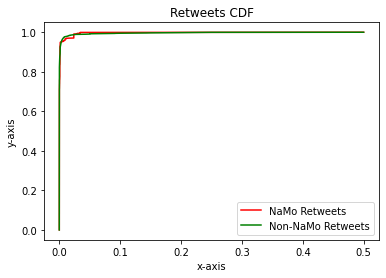}
        \subcaption{Plot for retweets}\label{fig:testa}
        \end{center}
    \end{subfigure}
    \begin{subfigure}[b]{0.49\textwidth}
        \begin{center}
        \includegraphics[width=60mm, height=50mm]{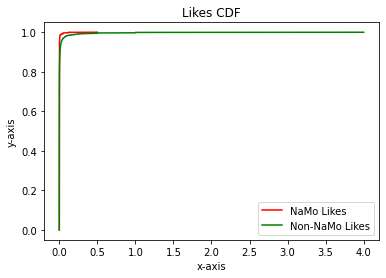}
        \subcaption{Plot for likes}\label{fig:testb}
        \end{center}
    \end{subfigure}
    \caption{CDF graphs for 2019 Lok Sabha Elections.}\label{}
\end{figure}

\begin{figure}[H]
    \begin{subfigure}[b]{0.49\textwidth}
        \begin{center}
        \includegraphics[width=60mm, height=50mm]{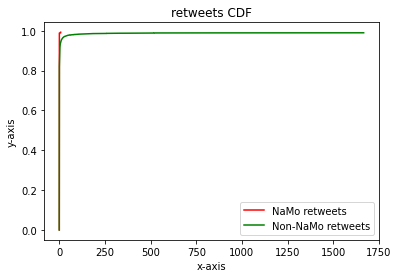}
        \subcaption{Plot for retweets}\label{fig:testa}
        \end{center}
    \end{subfigure}
    \begin{subfigure}[b]{0.49\textwidth}
        \begin{center}
        \includegraphics[width=60mm, height=50mm]{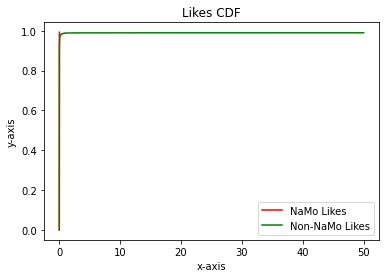}
        \subcaption{Plot for likes}\label{fig:testb}
        \end{center}
    \end{subfigure}
    \caption{CDF graphs for CAA Protests.}\label{}
\end{figure}

\subsection{Observations}
Through CDF and the correlation graphs for both the datasets, we observed that there are some outliers. Some Non-NaMo tweets have a very high number of likes and retweets than their corresponding NaMo tweet. However, we received the same result even when we normalized the parameters, i.e., likes and retweets, by user's followers count. Both NaMo content and its mapped Non-NaMo tweets have significantly less engagement on Twitter. When we compared the tweets posted using NaMo with its mapped Non-NaMo tweet, we observed that NaMo tweets do not receive as many retweets or likes as the posts that had the same content but posted without using the NaMo App (Non-NaMo tweets). Hence, we can infer that content posted through the NaMo App may not have much reachability and thus, may not affect the larger discourse on Twitter.

%% file: Chapters/Chapter10.tex

\section{Conclusion and Discussion} 

\label{Conclusion and Discussion} 

\subsection{Conclusion}
Through this research, we were able to conclude that:
\begin{enumerate}
\item{Due to the clustering of users of BJP and ‘Other Users’ at opposite ends of the plot
comparing consumption polarity and production polarity, these users tend to
consume and produce content that conforms to their ideology. Hence, there is an existence
of online echo chambers on Twitter.}
\item{In the graph of the follow network, the creation of various communities of left-leaning (anti-BJP and pro-Congress) and right-leaning (pro-BJP) users indicates that these various political leanings are transferred into social media networks in the form of closely linked groups. Also, partitioning indicates that active users predominately belongs to BJP-oriented ideology.}
\item{In comparison to the quantity of tweets in the database, the amount of content shared from the NaMo App to Twitter is less. Furthermore, tweets that are posted from the NaMo App are often already posted on Twitter by auxiliary users signifying less information flow.}
\item{In description of auxiliary users, N-grams such as ‘union minister', ‘member legislative
assembly' tend to have a higher odds ratio. This suggests to the possibility of association of auxiliary users with influential poeple in BJP party. Thus, they are posting the same content but not using NaMo App. In description of seed users, n-grams such as ‘{\dn aAIVF s\2yojk}', ‘{\dn EjlA \3FEw\7{m}K}', `{\dn aAIVF}', ‘social media soldiers' have higher odds ratio, indicating that posting on Twitter
using NaMo App are more likely to be done by BJP workers and supporters.}
\item{The reachability of the NaMo content on Twitter is very limited. Thus, the content from NaMo is not affecting the larger discourse and shows its inability to change the narrative on Twitter.Also, the users who posts content from the NaMo App are part of an online echo chamber, showing its inability to break into a wider and more diverse audience.}
\end{enumerate}

\subsection{Future work}
\begin{enumerate}
    \item{To identify the users that BJP should target to increase its reach or influence through NaMo App.}
    \item{A deeper content analysis on the basis of topics to understand if certain topics posted from NaMo are more influential on Twitter or more often cross posted.}
    \item{More deeper user analysis - Users present on NaMo and common users between both the platforms.}
\end{enumerate}

\subsection{Limitations and Challenges}
\textbf{Naive Political classification:} Due to the limitation of time in this thesis, this research was majorly based on the BJP vs. Anti-BJP/Pro-Congress. We classified the users in a binary class such as BJP or Others (Anti-BJP and Pro-Congress) based on our manually curated list of top 600 hashtags that trended during the 2019 Lok Sabha Election. Our dataset of top 600 hashtags contained more Pro-BJP hashtags in comparison to other hashtags. To understand the users and their political affiliation in-depth, the research can be extended into the multi-class problem by considering a reasonable number of hashtags associated with different parties. For example, a user can be classified based on several political leanings such as Pro-BJP, Anti-BJP, Pro-Cong, Anti-Cong, or other political parties. 
Also, our calculation of the users' polarity does not consider the tweet's text and depends entirely on the scores of hashtags. It may lead to assigning wrong political affiliation to users because the intent of the tweet may not align with the hashtags used in the tweet. For instance, a user can post a tweet: ‘Modi has done nothing for the country in his five years of constituency \#ModiAaneWalaHai \#ModiHaiToMumkinHai.' In this thesis, we assumed that most of the tweets are not sarcastic, although there is always a possibility of substantial occurrences of such tweets. 
\\

\textbf{Neutral Hashtags:} We did not consider the neutral hashtags while assigning the political association to users. Neutral hashtags are those hashtags which are not aligned to any political party such as \#2019LokSabhaElections, \#ExitPolls2019, \#VoteForIndia, etc. It may result in allocating a leaning to a user who has used more neutral hashtags than annotated hashtags even though the user is not that politically active. 
\\

\textbf{User Count Ratios:} We compared the ratios of the number of affected users by the total number of users (general users) for different states during the 2019 Lok Sabha Elections. The number of users for states may differ for different periods because users' tweeting activity may vary with geographically trending issues. To better understand the count of affected users for different states, one can compare them by incorporating a variety of tweets from events happening around the world and not limiting the dataset to just Lok Sabha/Elections datasets.
\\

\textbf{Code-mix Data:} A challenge that we faced during the thesis was handling code-mixed data. The datasets we had, consisted of tweets in various languages. The multilingual data made it challenging to remove stopwords to create word clouds and also made it difficult to give good results for algorithms such as LDA modeling. However, with future research and developments in Natural Language Processing, this issue can indeed be tackled.

%% file: ms.bbl
\begin{thebibliography}{10}

\bibitem{6}
2014 to 2019: From social media to app, bjp’s digital campaign shifts gear.
\newblock
  \url{https://indianexpress.com/article/india/2014-to-2019-social-media-namo-app-bjp-digital-campaign-5448706/},
  Nov. 2018.
\newblock accessed July 22, 2021.

\bibitem{30}
General election results social media reactions: As nda leads in over 300
  seats, supporters cheer online.
\newblock
  \url{https://indianexpress.com/article/trending/trending-in-india/election-2019-live-updates-5742921/},
  may 2019.
\newblock accessed July 22, 2021.

\bibitem{27}
Lok sabha election 2019: Full schedule with dates, phases, constituency-wise
  details; all you need to know.
\newblock
  \url{https://www.firstpost.com/india/lok-sabha-election-2019-full-schedule-with-dates-phases-constituency-wise-details-all-you-need-to-know-6411351.html},
  may 2019.
\newblock accessed July 22, 2021.

\bibitem{44}
{\sc Abbey, R.}
\newblock Cass r. sunstein. \#republic: Divided democracy in the age of social
  media . princeton, nj: Princeton university press, 2017. pp. xi+310. \$29.95.
\newblock {\em American Political Thought 7\/} (03 2018), 370--373.

\bibitem{50}
{\sc Agarwal, V., Vekaria, Y., Agarwal, P., Mahapatra, S., Set, S., Muthiah,
  S.~B., Sastry, N., and Kourtellis, N.}
\newblock Under the spotlight: Web tracking in indian partisan news websites,
  2021.

\bibitem{16}
{\sc An, J., Quercia, D., Cha, M., Gummadi, K.~P., and Crowcroft, J.}
\newblock Sharing political news: The balancing act of intimacy and
  socialization in selective exposure.
\newblock {\em EPJ Data Science 3\/} (12 2014), 1--21.

\bibitem{42}
{\sc Anuja, and Khanna, P.}
\newblock 2019 lok sabha election clocks highest ever turnout at 67.11\%.
\newblock
  \url{https://www.livemint.com/elections/lok-sabha-elections/at-67-11-2019-turnout-highest-for-lok-sabha-polls-1558376272609.html},
  may 2019.
\newblock accessed July 22, 2021.

\bibitem{7}
{\sc Bansal, S.}
\newblock Disfact \#18: How fake news thrives on the narendra modi app.
\newblock
  \url{https://bansalsamarth.substack.com/p/disfact-18-how-fake-news-thrives},
  Jan. 2019.
\newblock accessed July 22, 2021.

\bibitem{48}
{\sc Bhatt, S., Joglekar, S., Bano, S., and Sastry, N.}
\newblock Illuminating an ecosystem of partisan websites.
\newblock In {\em Companion Proceedings of the The Web Conference 2018\/}
  (Republic and Canton of Geneva, CHE, 2018), WWW '18, International World Wide
  Web Conferences Steering Committee, p.~545–554.

\bibitem{11}
{\sc Cohen, R., and Ruths, D.}
\newblock Classifying political orientation on twitter: It's not easy!
\newblock In {\em ICWSM\/} (2013).

\bibitem{9}
{\sc Conover, M.~D., Goncalves, B., Ratkiewicz, J., Flammini, A., and Menczer,
  F.}
\newblock Predicting the political alignment of twitter users.
\newblock In {\em 2011 IEEE Third International Conference on Privacy,
  Security, Risk and Trust and 2011 IEEE Third International Conference on
  Social Computing\/} (2011), pp.~192--199.

\bibitem{12}
{\sc Conover, M.~D., Ratkiewicz, J., Francisco, M., Gonçalves, B., Menczer,
  F., and Flammini, A.}
\newblock Political polarization on twitter.
\newblock In {\em ICWSM\/} (2011).

\bibitem{40}
{\sc Dale, B., and Jeavans, C.}
\newblock India general election 2019: What happened?
\newblock {\em BBC News\/} (may 2019).

\bibitem{13}
{\sc Du, S., and Gregory, S.}
\newblock The echo chamber effect in twitter: does community polarization
  increase?
\newblock In {\em Complex Networks {\&} Their Applications V\/} (Cham, 2017),
  H.~Cherifi, S.~Gaito, W.~Quattrociocchi, and A.~Sala, Eds., Springer
  International Publishing, pp.~373--378.

\bibitem{1}
{\sc Dutta, S., and Fraser, M.}
\newblock Barack obama and the facebook election.
\newblock
  \url{https://www.usnews.com/opinion/articles/2008/11/19/barack-obama-and-the-facebook-election},
  2008.
\newblock accessed July 22, 2021.

\bibitem{25}
{\sc Finkelstein, J., Zannettou, S., Bradlyn, B., and Blackburn, J.}
\newblock A quantitative approach to understanding online antisemitism.
\newblock {\em CoRR abs/1809.01644\/} (2018).

\bibitem{46}
{\sc Fletcher, R.}
\newblock The truth behind filter bubbles: Bursting some myths.
\newblock
  \url{https://reutersinstitute.politics.ox.ac.uk/risj-review/truth-behind-filter-bubbles-bursting-some-myths}.
\newblock accessed July 22, 2021.

\bibitem{15}
{\sc Garimella, K., De~Francisci~Morales, G., Gionis, A., and Mathioudakis, M.}
\newblock The effect of collective attention on controversial debates on social
  media.
\newblock In {\em Proceedings of the 2017 ACM on Web Science Conference\/} (New
  York, NY, USA, 2017), WebSci '17, Association for Computing Machinery,
  p.~43–52.

\bibitem{55}
{\sc Garimella, K., De~Francisci~Morales, G., Gionis, A., and Mathioudakis, M.}
\newblock Reducing controversy by connecting opposing views.
\newblock In {\em Proceedings of the Tenth ACM International Conference on Web
  Search and Data Mining\/} (New York, NY, USA, 2017), WSDM '17, Association
  for Computing Machinery, p.~81–90.

\bibitem{5}
{\sc Garimella, K., De~Francisci~Morales, G., Gionis, A., and Mathioudakis, M.}
\newblock Political discourse on social media: Echo chambers, gatekeepers, and
  the price of bipartisanship.
\newblock In {\em Proceedings of the 2018 World Wide Web Conference\/}
  (Republic and Canton of Geneva, CHE, 2018), WWW '18, International World Wide
  Web Conferences Steering Committee, p.~913–922.

\bibitem{14}
{\sc Garimella, K., and Weber, I.}
\newblock A long-term analysis of polarization on twitter.
\newblock {\em CoRR abs/1703.02769\/} (2017).

\bibitem{47}
{\sc Garrett, R.~K.}
\newblock Echo chambers online?: Politically motivated selective exposure among
  internet news users1.
\newblock {\em Journal of Computer-Mediated Communication 14}, 2 (2009),
  265--285.

\bibitem{49}
{\sc Garrett, R.~K., Long, J.~A., and Jeong, M.~S.}
\newblock {From Partisan Media to Misperception: Affective Polarization as
  Mediator}.
\newblock {\em Journal of Communication 69}, 5 (10 2019), 490--512.

\bibitem{38}
{\sc Ghosh, S.}
\newblock Elections 2019: India's youth is talking; will the political class
  listen?

\bibitem{58}
{\sc Gupta, P., and Shrimankar, D.}
\newblock How nationalism helped the bjp.
\newblock {\em Seminar 720\/} (2019), 38–43.

\bibitem{39}
{\sc Inc, G.}
\newblock 2014 vs 2019: A detailed general election results comparison with
  data.

\bibitem{54}
{\sc Jackson, S.}
\newblock {The Double-Edged Sword of Banning Extremists from Social Media}.
\newblock SocArXiv 2g7yd, Center for Open Science, July 2019.

\bibitem{43}
{\sc Katju, M.}
\newblock Election campaigning in a transformed india.
\newblock
  \url{https://www.theindiaforum.in/article/campaigning-transformed-india},
  january 2020.
\newblock accessed July 22, 2021.

\bibitem{52}
{\sc Kitchens, B., Johnson, S., and Gray, P.}
\newblock Understanding echo chambers and filter bubbles: The impact of social
  media on diversification and partisan shifts in news consumption.
\newblock {\em MIS Quarterly 44\/} (08 2020).

\bibitem{56}
{\sc Linderman, S., and Adams, R.}
\newblock Discovering latent network structure in point process data.
\newblock In {\em Proceedings of the 31st International Conference on Machine
  Learning\/} (Bejing, China, 22--24 Jun 2014), E.~P. Xing and T.~Jebara, Eds.,
  vol.~32 of {\em Proceedings of Machine Learning Research}, PMLR,
  pp.~1413--1421.

\bibitem{26}
{\sc Linderman, S.~W., and Adams, R.~P.}
\newblock Scalable bayesian inference for excitatory point process networks,
  2015.

\bibitem{8}
{\sc Manu, D., Krishnan, R., and Kumaraguru, P.}
\newblock Analysing how the shift in discourses on social media affected the
  narrative around the indian general election 2019.
\newblock {\em Journal of Advanced Research in Social Sciences 3\/}, 21--31.

\bibitem{4}
{\sc Mehta, N.}
\newblock Digital politics in india’s 2019 general elections, Dec 2019.

\bibitem{51}
{\sc Mishra, D., and Pal, J.}
\newblock The freedom of press and social media partisanship in india: A
  visualization of tweets around the 2020 fir against the wire.
\newblock
  \url{http://joyojeet.people.si.umich.edu/freedom-of-press-and-social-media-partisanship-in-india/}.
\newblock accessed July 22, 2021.

\bibitem{17}
{\sc Morales, G., Monti, C., and Starnini, M.}
\newblock No echo in the chambers of political interactions on reddit.
\newblock {\em Scientific Reports 11\/} (02 2021).

\bibitem{29}
{\sc Neha, K., Srikanth, S., Singhal, S., Singh, S., Buduru, A.~B., and
  Kumaraguru, P.}
\newblock Is change the only constant? profile change perspective on
  {\#}loksabhaelections2019.
\newblock {\em CoRR abs/1909.10012\/} (2019).

\bibitem{31}
{\sc Panda, A., Gonawela, A., Acharyya, S., Mishra, D., Mohapatra, M.,
  Chandrasekaran, R., and Pal, J.}
\newblock Nivaduck - a scalable pipeline to build a database of political
  twitter handles for india and the united states.
\newblock In {\em International Conference on Social Media and Society\/} (New
  York, NY, USA, 2020), SMSociety'20, Association for Computing Machinery,
  p.~200–209.

\bibitem{45}
{\sc Parser, E.}
\newblock The filter bubble how the new personalized web is changing what we
  read and how we think.
\newblock {\em Penguin Random House\/} (april 2012).

\bibitem{10}
{\sc Pennacchiotti, M., and Popescu, A.-M.}
\newblock Democrats, republicans and starbucks afficionados: User
  classification in twitter.
\newblock In {\em Proceedings of the 17th ACM SIGKDD International Conference
  on Knowledge Discovery and Data Mining\/} (New York, NY, USA, 2011), KDD '11,
  Association for Computing Machinery, p.~430–438.

\bibitem{36}
{\sc Prakash, A.}
\newblock The dynamics of social media and the indian elections 2019.
\newblock \url
  {https://theasiadialogue.com/2019/04/12/the-dynamics-of-social-media-and-the-indian-elections-2019/},
  april 2019.
\newblock accessed July 22, 2021.

\bibitem{3}
{\sc PTI}.
\newblock Social media plays key role in influencing first-time voters: Report.
\newblock
  \url{https://economictimes.indiatimes.com/news/elections/lok-sabha/india/social-media-plays-key-role-in-influencing-first-time-voters-report/articleshow/69295605.cms?from=mdr},
  may 2019.
\newblock accessed July 22, 2021.

\bibitem{2}
{\sc Rao, S.}
\newblock Making of selfie nationalism: Narendra modi, the paradigm shift to
  social media governance, and crisis of democracy.
\newblock {\em Journal of Communication Inquiry 42}, 2 (2018), 166--183.

\bibitem{37}
{\sc Sadashivam, T.}
\newblock Role of social media in indian general election 2019.
\newblock
  \url{https://www.academia.edu/38596908/Role_of_Social_Media_in_Indian_General_Election_2019},
  03 2019.
\newblock accessed July 22, 2021.

\bibitem{53}
{\sc Stasavage, D.}
\newblock Polarization and publicity: Rethinking the benefits of deliberative
  democracy.
\newblock {\em The Journal of Politics 69}, 1 (2007), 59--72.

\bibitem{35}
{\sc Stewart, A.~J., Mosleh, M., Diakonova, M., Arechar, A.~A., Rand, D.~G.,
  and Plotkin, J.~B.}
\newblock Information gerrymandering and undemocratic decisions.
\newblock {\em Nature 573\/} (september 2019), 117--121.

\bibitem{34}
{\sc Szumilas, M.}
\newblock Explaining odds ratios.
\newblock {\em Journal of the Canadian Academy of Child and Adolescent
  Psychiatry = Journal de l'Academie canadienne de psychiatrie de l'enfant et
  de l'adolescent 19}, 3 (August 2010), 227—229.

\bibitem{28}
{\sc Varol, O., Ferrara, E., Ogan, C.~L., Menczer, F., and Flammini, A.}
\newblock Evolution of online user behavior during a social upheaval.
\newblock In {\em Proceedings of the 2014 ACM Conference on Web Science\/} (New
  York, NY, USA, 2014), WebSci '14, Association for Computing Machinery,
  p.~81–90.

\bibitem{57}
{\sc Woetzel, D.}
\newblock Fake news: Separating fact from fiction.
\newblock \url{https://libguides.reynolds.edu/fakenews/bias}, 06 2021.
\newblock accessed July 22, 2021.

\bibitem{20}
{\sc Zannettou, S., Caulfield, T., Blackburn, J., De~Cristofaro, E.,
  Sirivianos, M., Stringhini, G., and Suarez-Tangil, G.}
\newblock On the origins of memes by means of fringe web communities.
\newblock In {\em Proceedings of the Internet Measurement Conference 2018\/}
  (New York, NY, USA, 2018), IMC '18, Association for Computing Machinery,
  p.~188–202.

\bibitem{21}
{\sc Zannettou, S., Caulfield, T., Bradlyn, B., Cristofaro, E.~D., Stringhini,
  G., and Blackburn, J.}
\newblock Characterizing the use of images in state-sponsored information
  warfare operations by russian trolls on twitter, 2019.

\bibitem{22}
{\sc Zannettou, S., Caulfield, T., Cristofaro, E.~D., Kourtellis, N.,
  Leontiadis, I., Sirivianos, M., Stringhini, G., and Blackburn, J.}
\newblock The web centipede: Understanding how web communities influence each
  other through the lens of mainstream and alternative news sources, 2017.

\bibitem{23}
{\sc Zannettou, S., Caulfield, T., Cristofaro, E.~D., Sirivianos, M.,
  Stringhini, G., and Blackburn, J.}
\newblock Disinformation warfare: Understanding state-sponsored trolls on
  twitter and their influence on the web.
\newblock {\em CoRR abs/1801.09288\/} (2018).

\bibitem{24}
{\sc Zannettou, S., Caulfield, T., Setzer, W., Sirivianos, M., Stringhini, G.,
  and Blackburn, J.}
\newblock Who let the trolls out? towards understanding state-sponsored trolls.
\newblock {\em CoRR abs/1811.03130\/} (2018).

\bibitem{33}
{\sc Zauner, C., Steinebach, M., and Hermann, E.}
\newblock {Rihamark: perceptual image hash benchmarking}.
\newblock In {\em Media Watermarking, Security, and Forensics III\/} (2011),
  N.~D. Memon, J.~Dittmann, A.~M. Alattar, and E.~J.~D. III, Eds., vol.~7880,
  International Society for Optics and Photonics, SPIE, pp.~343 -- 357.

\end{thebibliography}
